%% file: ms.tex
\documentclass{aastex}
\usepackage{subfigure}
\newcommand{\vectm}[1]{ \mbox{\boldmath $#1$}}

\shortauthors{Suetsugu, Ohtsuki, \&  Fujita}
\shorttitle{Orbital Characteristics of Captured Planetesimals}

\begin{document}

\title{Orbital Characteristics of Planetesimals Captured by Circumplanetary Gas Disks
}
\author{Ryo Suetsugu\altaffilmark{1,2}, Keiji Ohtsuki\altaffilmark{2},  
and Tetsuya Fujita\altaffilmark{2,3}}
\affil{1. Organization of Advanced Science and Technology, Kobe University, 
Kobe 657-8501, Japan}
\affil{2. Department of Planetology, Kobe University, 
Kobe 657-8501, Japan}
\affil{3. OGIS-RI Co., Ltd., Osaka 550-0023, Japan}
\email{suetsugu@buffalo.kobe-u.ac.jp, ohtsuki@tiger.kobe-u.ac.jp}

\begin{abstract}
Sufficiently massive growing giant planets have circumplanetary disks, and the capture of solid bodies by the disks would likely influence the growth of the planets and formation of satellite systems around them. 
In addition to dust particles that are supplied to the disk with inflowing gas, recent studies suggest the importance of capture of planetesimals whose motion is decoupled from the gas, but orbital evolution of captured bodies in circumplanetary gas disks has not been studied in detail. 
In the present work, using three-body orbital integration and analytic calculations, we examine orbital characteristics and subsequent dynamical evolution of planetesimals captured by gas drag from circumplanetary gas disks.
We find that the semi-major axes of the planet-centered orbits of planetesimals at the time of permanent capture are smaller than about one third of the planet's Hill radius in most cases. 
Typically, captured bodies rapidly spiral into the planet, and the rate of the orbital decay is faster for the retrograde orbits due to the strong headwind from the circumplanetary gas. 
When a planetesimal captured into a retrograde orbit suffers from sufficiently strong gas drag before spiraling into the planet, its orbit turns to the prograde direction at a radial location that can be explained using the Stokes number. 
We also find that those captured into certain types of orbits can survive for a long period of time even under gas drag both in the prograde and retrograde cases, which may be important for the origin of irregular satellites of giant planets.

\end{abstract}

\input{sec1}
\input{sec2}

\input{sec3}

\input{sec4}
\input{sec5}

\input{sec6}

\input{sec7}

\input{reference}
\input{figcap}

\end{document}

%% file: sec1.tex
\section{INTRODUCTION}

Giant planets grow by accreting gas from the surrounding nebula gas, and they have circumplanetary disks when they grow sufficiently massive.
Capture of solid bodies by the circumplanetary gas disks is important for the growth of the planets and the formation of satellite systems around them.
As for the building blocks of regular satellites, in addition to dust particles that are supplied to the disk with inflowing gas \citep[e.g.,][]{CW02}, capture of planetesimals whose motion is decoupled from the gas would also be important \citep{E09, F13}.
\citet{F13} obtained rates of capture of such large planetesimals by gas drag from circumplanetary gas disks.
Assuming axisymmetric circumplanetary gas disks, they performed analytic calculation and three-body orbital integration for planetesimals under gas drag, and found that a growing giant planet embedded in the protoplanetary disk with uniform radial distribution of planetesimals can capture planetesimals with sizes depending on the strength of gas drag.
More recently, \citet{T14} further investigated capture of planetesimals by taking account of results of hydrodynamic simulations of gas flow around a growing giant planet, and found that capture rates have a peak for planetesimals with a certain range of sizes.

While \citet{F13} focused on the capture processes of planetesimals by circumplanetary gas disks, orbital evolution of captured planetesimals is also important in studying various influences of captured bodies on the formation and evolution of satellite systems.  
For example, their orbital evolution in the disk would likely affect the radial distribution of solid bodies in the circumplanetary disk, which would be related to the location and timescale of satellite formation.
While three-body gravitational interaction has been proposed as a promising mechanism for capture of irregular satellites of giant planets \citep [e.g.,][]{N07, N14}, capture by gas drag from circumplanetary gas disks may also have played a role \citep[e.g.,][]{P79, CB04}.
Furthermore, collision of captured planetesimals may have influenced surface evolution of regular satellites \citep{B13}, and dust generated by collision between captured planetesimals could be observed around extrasolar planets \citep{KW11}.

Such influences of captured bodies on the planet and the satellite system largely depend on the radial location of capture,
which affects the strength of gas drag and lifetime of captured bodies in the disk. 
In their analytic calculation, \citet{F13} considered capture of planetesimals by a single encounter with the planet under relatively strong gas drag, and obtained a critical radial distance from the planet within which planetesimals are captured.
For example, large bodies under relatively weak gas drag can be captured only in the vicinity of the planet where the gas density in the disk is higher, and they spiral into the planet rather quickly after the capture due to the strong gas drag.
However, in the case of bodies captured into such orbits that keep a certain distance from the planet, they may stay in the disk for a longer period of time.
\citet{S11} and \citet{SO13}  examined temporary capture of planetesimals by a planet and found that there are four kinds of long-lived capture orbits, but they did not take account of gas drag.
Although those planetesimals that experience a significant period of temporary capture before becoming permanently captured may have been included in the calculations of \citet{F13}, they did not examine orbital evolution of captured planetesimals in detail. 

In the present work, as an extension of the work by \citet{F13}, we will examine orbital evolution of captured planetesimals in circumplanetary gas disks.
As in Fujita et al., we will focus on large planetesimals that are decoupled from the gas flow accreting onto the planet.
In Section~\ref{sec:model}, we describe our model and numerical methods.
In Section~\ref{sec:capture}, we examine distribution of orbital elements of planetesimals immediately after they become permanently captured.
We then examine orbital evolution of captured bodies by showing some examples in typical cases in Section~\ref{sec:example}.
Section~\ref{sec:long-lived} presents examples of long-lived orbits in the circumplanetary gas disk.
In Section~\ref{sec:orbevo_timescale}, we show that orbital evolution of captured planetesimals can be divided into a couple of stages in terms of their Stokes number, and examine timescales of orbital evolution.
Our conclusions are summarized in Section~\ref{sec:conclusion}.

%% file: sec2.tex
\section{THE MODEL}
\label{sec:model}

\subsection{Equations of Motion}

We consider the three-body problem for the Sun (mass $M_\odot$), a planet ($M$), and a planetesimal ($m_s$), and assume that the planet has a circumplanetary gas disk \citep{F13}.
In a local, rectangular and rotating coordinate system centered on the planet, the equations of motion are given by \citep{N89, O12}
  \begin{eqnarray}
     \ddot{x} &=&   2\Omega\dot{y} + 3\Omega^2x - \frac{G(M+m_s)}{R^3}x + a_{{\rm drag},x}, \nonumber \\
     \ddot{y} &=& - 2\Omega\dot{x} - \frac{G(M+m_s)}{R^3}y + a_{{\rm drag},y}, \label{eom-i}\\
     \ddot{z} &=& - \Omega^2z - \frac{G(M+m_s)}{R^3}z + a_{{\rm drag},z}. \nonumber
  \end{eqnarray}
\noindent
In the above, $\Omega$ is the planet's orbital angular frequency, and $R=\sqrt{x^2+y^2+z^2}$ is the distance between the centers of the planet and the planetesimal. 
$\vectm{a}_{\rm drag} \equiv {\vectm{F}}_{\rm drag}/m_s$ is the acceleration due to the gas drag force ${\vectm{F}}_{\rm drag}$ given by
  \begin{eqnarray}
	 {\vectm{F}}_{\rm drag} = \frac12 C_D \pi r_s^2 \rho_{\rm gas} u{\vectm{u}},
  \end{eqnarray}
where  $C_D$ is the drag coefficient (we assume $C_D=1$), $r_s$ is the radius of the planetesimal,  $\rho_{\rm gas}$ is the gas density, 
and $u=|{\vectm{u}}|$ is the velocity of the planetesimal relative to the gas.
We scale the distance by mutual Hill radius $r_H \equiv ah_H$ ($h_H = \{(M+m_s)/3M_\odot\}^{1/3}$ and $a$ is the semimajor axis of the planet) and time by $\Omega^{-1}$.
Then, we can express the above equation of motion in a non-dimensional form as
  \begin{eqnarray} \label{eq:hilleq}
    && \ddot{\tilde{x}} =   2\dot{\tilde{y}} + 3\tilde{x} - \frac{3\tilde{x}}{\tilde{R}^3} + \tilde{a}_{{\rm drag},x} \nonumber \\
    && \ddot{\tilde{y}} = - 2\dot{\tilde{x}} - \frac{3\tilde{y}}{\tilde{R}^3} + \tilde{a}_{{\rm drag},y} \label{eom}\\
    && \ddot{\tilde{z}} = - \tilde{z} - \frac{3\tilde{z}}{\tilde{R}^3} + \tilde{a}_{{\rm drag},z} \nonumber \label{eq:nondimeq}
  \end{eqnarray}
where tildes denote non-dimensional quantities \citep{TO10, F13}.
The non-dimensional acceleration due to gas drag, $\tilde{\vectm{a}}_{\rm drag}$, can be described as
  \begin{equation} \label{eq:drag-term1}
    \tilde{\vectm{a}}_{\rm drag} \equiv \frac{{\vectm{F}}_{\rm drag}/m_s}{r_H\Omega^2} = -\frac{3}{8}C_{\rm D} \frac{\rho_{\rm gas}}{\tilde{r}_s\rho_s} \tilde{u} \tilde{\vectm{u}},
  \end{equation}
where $\rho_s$ is the internal density of the planetesimal.

\subsection{Disk Structure}
Results of recent high-resolution hydrodynamic simulations show that the circumplanetary disk can be approximated to be axisymmetric in the vicinity of the planet \citep[e.g.,][]{M08, T12}.
In the present work, we assume an axisymmetric thin circumplanetary disk, as in \citet{F13}.
We also assume that the radial distribution of the gas density is given by a power law,
and that the disk is vertically isothermal.
Under these assumptions, the gas density can be written as
    \begin{equation} \label{eq:rho_gas}
      \rho_{\rm gas} = \frac{\Sigma}{\sqrt{2\pi}h}{\rm exp}\left(-\frac{z^2}{2h^2}\right),
    \end{equation}
where $h = c_s/\Omega_K$ is the scale hight of the circumplanetary disk ($\Omega_K$ is the Keplerian orbital angular frequency around the planet), and 
	\begin{eqnarray} \label{eq:sigma,cs}
		\Sigma = \Sigma_d \left(\frac{r}{r_d}\right)^{-p}, \quad c_s = c_d \left(\frac{r}{r_d}\right)^{-q/2}
	\end{eqnarray}
are the gas surface density and the sound velocity, respectively, with $r=\sqrt{x^2+y^2}$ being the horizontal distance from the planet in the mid-plane.
In the above, $r_d \equiv dr_H$ is a typical length scale roughly corresponding to the effective size of the circumplanetary disk, and $c_d$ is the sound speed there.
We set $d=0.2$, and $p=3/2$ based on results of hydrodynamic simulation \citep{M08, T12}, and also assume $q=1/2$ \citep{F13}. 
The gas velocity can be written as
	\begin{eqnarray} %()
		v_{\rm gas} = (1-\eta)v_K,
	\end{eqnarray}
where $v_K$ is the Keplerian velocity around the planet, 
and $\eta$ is a small quantity that depends on the radial pressure gradient; using Equations (\ref{eq:rho_gas}) and (\ref{eq:sigma,cs}), 
$\eta$ can be written as \citep{T02}
	\begin{eqnarray} \label{eq:eta}%()
		\eta \equiv \frac{1}{2}\frac{h^2}{r^2}\left(p + \frac{q+3}{2} + \frac{q}{2}\frac{z^2}{h^2}\right).		
	\end{eqnarray}
Although we have defined the effective size of the circumplanetary disk $r_d$ in the above, we turn on gas drag when planetesimals enter within the planet's Hill sphere in order to avoid effects of artificial cutoff at $r=r_d$.
Since $\rho_{\rm gas}$ decreases sharply with increasing $r$, gas drag is significant only in the inner part of the disk and negligible in the outer region.

\subsection{Gas Drag Parameter $\zeta$}
When the gas density is given by  Equation (\ref{eq:rho_gas}), Equation (\ref{eq:drag-term1}) can be rewritten as \citep{F13}
  \begin{equation} \label{drag-term2}
    \tilde{\vectm{a}}_{\rm drag} = -\zeta \tilde{r}^{-\gamma}{\rm exp}\left(-\frac{\tilde{z}^2}{2\tilde{h}^2}\right) \tilde{u} \tilde{\vectm{u}},
  \end{equation}
where  $h=h_d(r/r_d)^{(3-q)/2}=h_d(r/r_d)^{5/4}$ ($h_d$ is the scale height at $r=r_d$), $\gamma \equiv p+(3-q)/2 = 11/4$, and $\zeta$ is the non-dimensional parameter representing the strength of gas drag defined by\footnote{Note that the value of $\Sigma_d=1$gcm$^{-2}$ gives $\Sigma\simeq10^{2}$gcm$^{-2}$ at $r\simeq10^{-2}r_H$, which is within the range of the surface density values at the radial location of the Galilean satellites in the gas-starved disk model \citep{CW02}.} 
  \begin{eqnarray} \label{eq:zeta}
    \zeta &\equiv& \frac{3}{8\sqrt{2\pi}} \frac{C_D}{r_s \rho_s} \frac{\Sigma_d}{\tilde{h}_d} d^\gamma \\
		  &=& 3 \times 10^{-7} C_D \left( \frac{r_s}{1 {\rm km}} \right)^{-1} \left( \frac{\rho_s}{10^3 {\rm kg\, m^{-3}}} \right)^{-1} \left( \frac{\Sigma_d}{1 {\rm g\, cm^{-2}}} \right) \left( \frac{\tilde{h}_d}{0.06} \right)^{-1} \left( \frac{d}{0.2} \right)^\gamma. \nonumber
  \end{eqnarray}
We set $\tilde{h}_d=0.06$ in the present work \citep{T12, F13}.
The functional form of $\zeta$ is similar to the inverse of the Stokes number;
the Stokes number at $r=r_d$ in the disk mid-plane can be written as 
  \begin{eqnarray} \label{eq:stokes}
	{\rm St} &=& \frac{u/a_{\rm drag}}{\Omega_K^{-1}} = \frac{8\sqrt{2\pi}}{3} \frac{r_s \rho_s}{C_D} \frac{h_d}{\Sigma_d} \frac{\Omega_K}{u} \\
	   &=& 7.8\times10^5 C_D^{-1} \left( \frac{r_s}{1 {\rm km}} \right) \left( \frac{\rho_s}{10^3 {\rm kg\, m^{-3}}} \right) \left( \frac{\Sigma_d}{1 {\rm g\, cm^{-2}}} \right)^{-1} \left( \frac{\tilde{h}_d}{0.06} \right) \tilde{u}^{-1}. \nonumber
  \end{eqnarray}
In this paper, we examine capture processes and orbital evolution of planetesimals that are large enough to be decoupled from the inflowing gas.
Thus, we will consider the case of ${\rm St} \gg 1$ at $r=r_d$ (i.e.,  $\zeta \ll 1$).
When $C_D$ and $\rho_s$ are given, $\zeta$ can be determined as a function of the gas surface density of the circumplanetary disk at $r=r_d$ and the planetesimal size $r_s$ \citep{F13}.

\subsection{Energy Dissipation Due to Gas Drag and Capture Radius}

We perform orbital integration of planetesimals with various initial orbital elements by numerically solving 
Equation~(\ref{eq:hilleq}), using the eighth-order Runge-Kutta integrator \citep{O93, OI98}.
Initial heliocentric orbits of planetesimals are defined by the eccentricity ($e$), inclination ($i$), semimajor axis difference between the planet and  a planetesimal ($b$), and horizontal and vertical phase angles $\tau$ and $\omega$, respectively (see \citet{F13} for a detailed description of numerical methods).
The initial energy of a planetesimal on the Hill coordinate system is given by \citep{N89}
	\begin{eqnarray} 
		\tilde{E} &=& \frac12(\dot{\tilde{x}}^2+\dot{\tilde{y}}^2+\dot{\tilde{z}}^2) - \frac12(3\tilde{x}^2-\tilde{z}^2)-\frac{3}{\tilde{R}} + \frac92 \nonumber \\
				  &=& \frac12 (e_{\rm H}^2+i_{\rm H}^2) - \frac38 b_{\rm H}^2 - \frac{3}{\tilde{R}} + \frac92 \label{eq:energy}
	\end{eqnarray}
where $e_{\rm H} \equiv e/h_{\rm H}$, $i_{\rm H} \equiv i/h_{\rm H}$, and $b_{\rm H} \equiv b/r_{\rm H}$.
The energy of planetesimals decreases by gas drag from the circumplanetary disk, and they become captured when the energy becomes negative within the planet's Hill sphere.

The amount of energy dissipation due to gas drag during the closest approach to the planet ($\Delta \tilde{E}$) can be approximately written as \citep{TO10, F13}  
	\begin{eqnarray} \label{eq:delE}
		\Delta \tilde{E} \sim \tilde{a}_{\rm drag} \tilde{l},
	\end{eqnarray}
where $\tilde{l}$ ($\sim f\tilde{r}_{\rm min}$; $f$ is a correction factor of order unity) is the path length of the orbit near the point of the closest approach. 
If planetesimals' random velocity is large enough to neglect gravitational interaction with the planet, 
their velocity  relative to the planet when they pass through the vicinity of the planet can be approximately given by \citep{N89}
$\tilde{v}_\infty = \sqrt{\tilde{e}^2 + \tilde{i}^2 - (3/4)\tilde{b}^2}$
, and $\tilde{v}_\infty = \sqrt{\tilde{e}^2-(3/4)\tilde{b}^2}$ in the coplanar case ($i_{\rm H}=0$).
On the other hand, when the random velocity is small, the orbit in the vicinity of the planet can be approximated by a parabola.

Using $\Delta \tilde{E}$ and $\tilde{v}_\infty$, \citet{F13} obtained the following analytic expressions for the amount of energy dissipation:
	\begin{eqnarray}
	\label{eq:E-dis_pro}
\Delta \tilde{E}_{\rm pro} \sim & 3f_{\rm pro} \zeta \tilde{r}_{\rm min}^{-\gamma} (3-2\sqrt2) 	& \quad \mbox{ for  prograde, shear-dominated case} \\
\label{eq:E-dis_retro}
\Delta \tilde{E}_{\rm retro} \sim & 3f_{\rm retro} \zeta \tilde{r}_{\rm min}^{-\gamma} (3+2\sqrt2) & \quad \mbox{ for retrograde, shear-dominated case} \\
\label{eq:E-dis_ran}
\Delta \tilde{E}_{\rm ran} \sim & f_{\rm ran}\zeta \tilde{r}_{\rm min}^{1-\gamma}\tilde{v}_\infty^2 & \quad \mbox{ for dispersion-dominated case}
	\end{eqnarray}
where $f_{\rm pro}, f_{\rm retro}$, and $f_{\rm ran}$ are correction factors of order unity. 
\citet{F13} defined the capture radius by the distance from the planet where the energy dissipation $\Delta \tilde{E}$ equals the initial energy $\tilde{E}$ of a planetesimal \citep[see also][]{TO10},
and obtained expressions for the capture radii for the above three cases:
	\begin{eqnarray}
	 \label{Rpro}
		\tilde{R}_{\rm shear, pro}   &\simeq& \left[ \frac{6(3-2\sqrt2)f_{\rm pro} \zeta}{\tilde{v}_\infty^2+9} \right]^{1/\gamma} \\
	 \label{Rretro}
		\tilde{R}_{\rm shear, retro} &\simeq& \left[ \frac{6(3+2\sqrt2)f_{\rm retro} \zeta}{\tilde{v}_\infty^2+9} \right]^{1/\gamma} \\
	 \label{Rran}
		\tilde{R}_{\rm ran}   &\simeq& \left[ \frac{2f_{\rm ran} \zeta \tilde{v}_\infty^2}{\tilde{v}_\infty^2+9} \right]^{1/(\gamma-1)}
	\end{eqnarray}
By combining these results, Fujita et al. obtained analytic expressions for the capture radii for the prograde and retrograde cases as
	\begin{eqnarray}
	\label{eq:pro-radi}
		\tilde{R}_{\rm pro} &=& \sqrt{\tilde{R}_{\rm ran}^2 + \tilde{R}_{\rm shear, pro}^2} \\
	\label{eq:ret-radi}
		\tilde{R}_{\rm retro} &=& \sqrt{\tilde{R}_{\rm ran}^2 + \tilde{R}_{\rm shear, retro}^2}
 	\end{eqnarray}
 	
\noindent

%% file: sec3.tex
\section{ORBITAL ELEMENTS OF PLANETESIMALS IMMEDIATELY AFTER PERMANENTLY CAPTURED}
\label{sec:capture}

Here we examine orbital elements of planet-centered orbits of planetesimals immediately after becoming permanently captured by gas drag from the circumplanetary disk.
First, we focus on the case of planetesimals that are captured via a single encounter with the disk due to strong gas drag.
Figure~\ref{fig:rmin_all} shows various quantities at the time of capture as functions of the minimum approach distance to the planet $\tilde{r}_{\rm min}$; total amount of energy dissipated before capture ($\Delta \tilde{E}$), semi-major axis of planet-centered orbits scaled by the planet's Hill radius ($a_{\rm p}/r_{\rm H}$) and their eccentricity ($e_{\rm p}$).
In the cases shown here, planetesimals have zero orbital inclinations initially (i.e., $i_{\rm H} = 0$), and cases with two different values of the gas drag parameter are shown in each panel.
These values of $\zeta$ roughly correspond to planetesimals with sizes 1m and 1km, respectively, if we assume $\Sigma_d=1$gcm$^{-2}$ in Equation~(\ref{eq:zeta}).
The amount of dissipated energy shown in the upper panels can also be regarded as the amount of initial energy $\tilde{E}$, because we are focusing on orbits immediately after permanently captured (i.e., $\tilde{E}$ is negative but very close to zero at the time of permanent capture).
For a given pair of orbital eccentricity and inclination, the initial energy is given by Equation (\ref{eq:energy}), which shows that those orbits with large values of $\Delta \tilde{E}$ at the time of capture (i.e., those with large initial values of $\tilde{E}$) represent orbits with small initial $b_{\rm H}$'s.
Capture takes place in a certain region of $\Delta \tilde{E}$ and $\tilde{r}_{\rm min}$, and the values of $\tilde{r}_{\rm min}$ for the outer boundary of each region represents the actual capture radius. 
For example, in the case of prograde capture (red points) for the case of $e_{\rm H} = 0.5$ (upper left panel), we can estimate the capture radius as $\tilde{R}_{\rm pro} \simeq 0.02-0.04$ in the case of $\zeta = 10^{-4}$, and $\tilde{R}_{\rm pro} \simeq 0.003 - 0.005$ in the case of $\zeta = 3 \times 10^{-7}$; planetesimals with weak gas drag (e.g., large planetesimals for a given gas density) can become captured only in the vicinity of the planet, where gas density is sufficiently high.
We find that the capture radius for the prograde orbits obtained numerically in this way roughly agree with the analytic estimates, while the numerically estimated capture radius in the retrograde case is somewhat smaller than the analytic ones.
The semi-major axes of planet-centered orbits at the time of capture in the case of $e_{\rm H} = 0.5$ distribute in a rather narrow region with $\tilde{a}_{\rm p} \simeq 0.3-0.35$, while the distribution extends further into small values in the case of $e_{\rm H} = 5$.
In most cases, orbital eccentricities at the time of capture are close to unity.

As shown in Figure~\ref{fig:rmin_all}, the distribution of orbital elements varies depending on the gas drag parameter and the initial eccentricity of heliocentric orbits.
However, one feature in common in all the cases shown here is that there is a maximum value of $\tilde{a}_{\rm p} \simeq 0.35$ at the time of capture.
This value can be analytically obtained as follows.
In typical cases, planetesimals become captured when they enter the inner region of the disk and experience sufficient energy dissipation.
In this case, owing to the proximity to the planet, we can neglect the tidal potential and the effect of the rotation of the coordinate system, and adopt the two-body approximation. 
Then, the sum of the kinetic energy and the mutual gravitational potential in Equation (\ref{eq:energy}) can be replaced by $-3/(2\tilde{a}_{\rm p})$, having $\tilde{E} \simeq -3/(2\tilde{a}_{\rm p})+(9/2)$.
Using this expression, the condition for capture, $\tilde{E} < 0$, can be rewritten as $\tilde{a}_{\rm p} < 1/3$.
This shows that the semi-major axes of captured planetesimals satisfy $a_{\rm p} < r_{\rm H}/3$ if the capture is caused by energy dissipation in the vicinity of the planet.
In the above analytic estimate we neglected the effects of tidal force and the rotation of the coordinate system, which becomes non-negligible for capture at radial locations close to the Hill radius.
Thus, the above analytic estimate is only an approximate one, but it seems to explain our numerical results quite well.

On the other hand, Figure~\ref{fig:rmin_long} shows the case of capture after multiple encounters with the planet. 
Orbits in this case typically show rather chaotic behavior.
Also, energy dissipation due to gas drag during such multiple encounters facilitates capture at distant locations where the gas density is lower.
As a result, capture in this case often takes place exterior to the analytically-derived capture radius both in the prograde and retrograde cases.
Such orbits with chaotic behavior becomes less common in the dispersion-dominated velocity regime \citep{IO07, S11}, showing better agreement with analytic results (the case with $e_{\rm H} = 5$; Figure~\ref{fig:rmin_long}(b)).
The direction of orbital motion (prograde or retrograde) of captured bodies is determined by their orbital angular momentum about the planet at the time of capture, thus it does not necessarily reflect the orbital direction in the preceding stage, such as the phase of temporary capture \citep{S11}. 
In some cases, the direction of orbital motion about the planet changes from retrograde to prograde due to the effect of gas drag; this will be discussed in detail using the Stokes number in Section~\ref{subsec:orb_evo}. 
We also notice in Figure~\ref{fig:rmin_long}(a) that, in the case of relatively weak gas drag with $\zeta=3\times10^{-7}$, there are a significant number of retrograde capture orbits with large $\tilde{r}_{\rm min}$ ($\gtrsim 0.02$) at $\Delta \tilde{E} = 2-4$. 
Planetesimals on these orbits become captured without passing through the dense part of the circumplanetary gas disk, and some of them stay within the disk for a long time. 
We will examine such long-lived orbits in Section~\ref{sec:long-lived}.

%% file: sec4.tex
\section{EXAMPLES OF ORBITAL BEHAVIOR OF CAPTURED PLANETESIMALS}
\label{sec:example}
Figures~\ref{fig:orb_pro} and \ref{fig:orb_ret} show orbital behavior and time variation of physical quantities for the case where capture takes place by strong gas drag at a single encounter with the circumplanetary disk.
Collision with the central planet is not taken into account in order to focus on orbital evolution due to gas drag.
Figure~\ref{fig:orb_pro} shows the case of capture in the prograde direction.
Eccentricities of planet-centered orbits immediately after the capture are rather large, and the radial distance from the planet oscillates in a wide range of $0.01 \lesssim \tilde{R} \lesssim 0.1$.
Every time the planetesimal passes the pericenter, it undergoes strong gas drag that results in rapid decrease in kinetic energy and semi-major axis, while keeping the pericenter distance nearly unchanged.
The eccentricity also decreases due to gas drag, and the orbit is circularized rather quickly.
With decreasing eccentricity, the relative velocity between the planetesimal and the surrounding gas also decreases; thus the rates of decrease in energy and semi-major axis are reduced at $e_{\rm p} \lesssim 0.2$.
Afterwards, eccentricities further decrease first, followed by gradual decrease in semi-major axes \citep{A76}.

Figure~\ref{fig:orb_ret} shows a typical case of capture in the retrograde direction.
The strong headwind causes capture of planetesimals and their rapid orbital decay in the circumplanetary disk.
Every time the planetesimal passes the pericenter, the eccentricity also decrease significantly.
When $e_{\rm p} \lesssim 0.2$, the apocenter becomes sufficiently close to the planet for the planetesimal to suffer from strong gas drag for the whole orbit, thus it spirals into the planet in a short timescale.
Figure~\ref{fig:orb_ret_up} shows a blow-up of Figure~\ref{fig:orb_ret}(b), where we can see that the direction of orbital motion is changed due to the strong headwind.
The change of the orbital direction can be confirmed from the change of the orbital angular momentum (Figure~\ref{fig:orb_ret}(c)), and the rapid increase of the orbital eccentricity at $\tilde{t} \simeq 4.81$ corresponds to this sudden turn. 
We will discuss the radial distance for such a change of orbital direction in Section~\ref{sec:orbevo_timescale}.

There are many other types of capture orbits, especially in the three-dimensional case with non-zero orbital inclinations.
In the three-dimensional case, capture by gas drag becomes difficult, because planetesimals penetrate the disk nearly vertically and the duration of interaction with the gas disk becomes short \citep{F13}.
Orbital inclinations of captured bodies decrease due to gas drag.
In the above, we have shown an example of the change of the orbital direction from a retrograde orbit to a prograde one in the mid-plane of the circumplanetary disk.
Figure~\ref{fig:ret_pro} shows a case of the change of orbital direction due to gas drag for a body on an off-plane orbit \citep[see also][]{CB04}.
In this case, the body undergoes strong gas drag every time it penetrates the inner dense part of the disk, which causes gradual decrease of the orbital inclination.
The distribution of $\tilde{r}_{\rm min}$ for captured planetesimals in the three-dimensional case tends to extend to the inner part of the disk, because stronger gas drag is required for planetesimals penetrating the disk nearly vertically than the coplanar case. 

%% file: sec5.tex
\section{LONG-LIVED PLANETOCENTERED ORBITS}
\label{sec:long-lived}

In the case of typical capture orbits shown in Section~\ref{sec:example}, captured planetesimals spiral into the planet in a short timescale ($\lesssim 10T_{\rm pla}$, where $T_{\rm pla}$ is the planet's orbital period).
However, there are other types of orbits that allow captured bodies to orbit about the planet for a longer time.
\citet{S11} examined long-lived temporary capture of planetesimal by a planet's gravity in the gas-free environment \citep[see also][]{SO13}.
\citet{CB04} discussed a possibility that a body that delivered into a long-lived temporary capture orbit under gas drag becomes a progenitor for a group of prograde irregular satellites of Jupiter.
Here we examine long-lived capture orbits under gas drag.

Long-lived temporary capture orbits in the prograde direction appear at very low energy ($\tilde{E}\simeq 0$). 
In this case, planetesimals enter the planet's Hill sphere through the vicinity of the Lagrangian points, and then they bounce back many times at the equipotential surface near the Hill sphere before escaping from it. 
Since the shape of the region swept by the trajectories become similar to the shape of the Hill sphere, 
\citet{S11} called this group of temporary capture orbits type-H orbits.
This type of orbits appear for a narrow range of eccentricities of the initial heliocentric orbits with $e_{\rm H} \simeq 3$.
Figure~\ref{fig:rmin_pro} shows distributions of the amount of dissipated energy and orbital elements at the time of permanent capture for such a case.
We find that there are orbits captured at distant regions with $\tilde{r}_{\rm min} \sim 10^{-2}-10^{-1}$ and $\Delta \tilde{E} \sim 10^{-3}-10^{-1}$ after multiple encounters with the planet, and we confirmed that these are type-H orbits.
In this case, the planet-centered orbits have larger semi-major axes and somewhat smaller eccentricities than the typical case of capture via a single encounter (Figures~\ref{fig:rmin_long}(b) and \ref{fig:rmin_long}(c)).
Figure~\ref{fig:orb_pro_long} shows orbital evolution and time variation of some quantities for this type of orbit.
Since the energy is close to zero, the initial orbital behavior is similar to the type-H temporary capture orbits in the gas-free environment.
Then, gas drag reduces the energy and the orbit gradually shrinks. 
An important characteristic of this type of orbits is that their pericenter distance is kept larger compared to the case of typical capture by a single encounter.
This allows captured bodies to avoid entering the dense part of the circumplanetary disk and survive in the disk for a longer time.

Figure~\ref{fig:orb_ret_long} shows an example of long-lived orbit in the retrograde direction.
The temporary capture orbit with an energy corresponding to this case is called type A, because the orbital shape is similar to the cross-section of an apple \citep{S11}.
The initial energy of the orbit shown in Figure~\ref{fig:orb_ret_long} is $\tilde{E} = 3.714$, close to the range of the energy for the type-A temporary capture orbits \citep[$\tilde{E} \simeq 2-3.5$;][]{S11}.
We find that those orbits with initial energy somewhat larger than the above range can also become this type of long-lived orbits owing to energy dissipation by gas drag.
Figure~\ref{fig:orb_ret_long} shows that planetesimals on this type of orbits can survive for a long time because they avoid entering the inner part of the disk, as in the case of prograde long-lived orbits.

Although the lifetimes of the above prograde and retrograde orbits are similar ($\gtrsim 10^3T_{\rm pla}$), their dynamical evolutions are somewhat different.
In the case of the prograde long-lived orbits (Figure~\ref{fig:orb_pro_long}), the planetesimal becomes permanently captured rather quickly because the initial energy is rather small, and the subsequent evolution before spiraling into the planet is slow.
On the other hand, in the retrograde case, the larger initial energy requires a longer time of interaction with the gas before the planetesimal becomes permanently captured.
The pericenter distance is kept rather large ($\tilde{R}>0.1$) during the phase of temporary capture, which results in slow decrease of the energy.
In fact, the orbit projected on the $\tilde{x}$-$\tilde{y}$ plane at $t/T_{\rm pla} = 1800$ seems to stay within the planet's Hill sphere, but the planetesimal still has a positive energy at this point and is not permanently captured yet. The decrease in orbital eccentricity and semi-major axis is also slow.
Once the body becomes permanently captured, the orbital decay proceeds quickly owing to the strong headwind.
In the three-dimensional case shown in Figure~\ref{fig:orb_ret_long}, the decrease of the orbital inclination is rather slow during the evolution.
The sudden increase in the eccentricity at the final stage corresponds to the change of the orbital direction that we mentioned before.

Figures~\ref{fig:orb_pro_long} and \ref{fig:orb_ret_long} show that planetesimals can orbit about a planet for a long time ($\sim 10^2-10^3 T_{\rm pla}$, which corresponds to $\sim 10^3-10^4$ years in the case of capture by Jupiter) under gas drag depending on the parameters.  
This orbital lifetime can become still longer if gradual dispersal of the circumplanetary gas disk is taken into account.
In this case, some of planetesimals captured on the long-lived orbits in the gas disk could survive as irregular satellites.
However, the possibility of survival most likely depends on the timing of the disk dispersal. 
We will investigate such an issue in detail in our separate work, where we take account of gradual dispersal of the circumplanetary gas disk \citep{SO16}.

%% file: sec6.tex
\section{ORBITAL EVOLUTION IN CIRCUMPLANETARY DISKS}
\label{sec:orbevo_timescale}

\subsection{Orbital Evolution and Stokes Number}
\label{subsec:orb_evo}

As we have shown in the previous sections (e.g., Figure~\ref{fig:orb_ret_up}), planetesimals captured into retrograde orbits can change their orbital direction as approaching the planet.
Here we analytically examine such orbital behavior using the Stokes number.

In terms of non-dimensional quantities, the Stokes number given by Equation (\ref{eq:stokes}) can be rewritten as 
\begin{eqnarray} \label{eq:stokes3} %()
		{\rm St} &=& \frac{8\sqrt{2\pi}}{3} \frac{r_s \rho_s}{C_D} \frac{h}{\Sigma} {\rm exp}\left( \frac{z^2}{2h^2} \right) \frac{\Omega_K}{u} \\ \nonumber
		   &=& \zeta^{-1} \tilde{r}^\gamma {\rm exp}\left( \frac{\tilde{z}^2}{2\tilde{h}^2} \right) \sqrt{\frac{3}{\tilde{r}^3}} \frac{1}{\tilde{u}}.
	\end{eqnarray}
In the two-dimensional case, this can be simplified as
	\begin{eqnarray} \label{eq:stokes2}%()
		{\rm St} = \zeta^{-1} \tilde{r}^\gamma \sqrt{\frac{3}{\tilde{r}^3}} \frac{1}{\tilde{u}}.
	\end{eqnarray}
In order to analytically examine orbital evolution after capture, we assume that the orbital eccentricities of captured bodies become negligibly small quickly due to gas drag both in the prograde and retrograde cases.
In this case, relative velocity between the body and the gas is given as $\tilde{u} \simeq \eta \tilde{v}_{\rm K}$ for the prograde case, and $\tilde{u} \simeq 2\tilde{v}_{\rm K}$ for the retrograde case. 
We have neglected the term arising from the pressure gradient in the gas disk in the retrograde case, since $\tilde{v}_{\rm K} \gg \eta \tilde{v}_{\rm K}$.
Using this approximation and Equation (\ref{eq:stokes2}), the Stokes number in the prograde and retrograde cases can be respectively given as 
 	\begin{eqnarray}
		{\rm St}_{\rm pro}   &=& \tilde{r}^{\gamma-1}/(\eta\zeta) \label{eq:stokes-pro}, \\
		{\rm St}_{\rm retro} &=& \tilde{r}^{\gamma-1}/(2\zeta) \label{eq:stokes-retro}.
	\end{eqnarray}
\noindent
As we can see from Equation (\ref{eq:eta}), $\eta$ can be given as a function of the radial distance from the planet alone in the two-dimensional case.
Thus, the Stokes number can also be analytically expressed as a function of $\tilde{r}$ alone under the above approximations if the disk structure is assumed.
On the other hand, the actual evolution of the Stokes number for each orbit can be calculated during our orbital integration.

Figure~\ref{fig:r_st} shows the plots of the variation of the Stokes number for three orbits as a function of the radial distance, together with the analytic results.
Figure~\ref{fig:r_st}(a) represents the case for the coplanar prograde capture orbit shown in Figures~\ref{fig:orb_pro}(a) and \ref{fig:orb_pro}(b).
The diagonal straight dotted line represents the analytic result for ${\rm St}_{\rm pro}$ given by Equation~(\ref{eq:stokes-pro}), and the horizontal dashed line shows  ${\rm St} = 1$.
The evolution can be divided into three stages.
Initially, the planetesimal is not permanently captured yet (blue curve), then becomes permanently captured (at the point shown by the open circle).
At this time the orbit is highly elongated, with its eccentricity being close to unity (Figure~\ref{fig:orb_pro}(c)).
Then, the eccentricity and the semi-major axis decrease rather rapidly due to gas drag (the part of the red curve with $\tilde{r} \gtrsim 10^{-2}$ in Figure~\ref{fig:r_st}(a)).
During this stage, the pericenter distance is kept nearly constant, at a value roughly corresponding to the capture radius for prograde orbits, $\tilde{R}_{\rm pro}$, given by Equation (\ref{eq:pro-radi}) (shown by the vertical dotted line).
Since the relative velocity between the planetesimal and the gas gradually decreases with decreasing eccentricity, the Stokes number gradually increases.
The sharp increase of the Stokes number ($> 10^5$ at $\tilde{r} \simeq 0.02$) is caused by an extremely low relative velocity with the gas at the apocenter.
After the eccentricity becomes sufficiently small ($\sim 0.01$), the reduced relative velocity decreases the rate of the orbital decay.
The variation of the Stokes number during this last stage of slow orbital decay follows the analytic result very well.

Figure~\ref{fig:r_st}(b) shows similar plots for the retrograde orbit shown in Figures~\ref{fig:orb_ret}(a) and \ref{fig:orb_ret}(b).
The planetesimal becomes permanently captured at a radial location close to the analytically-obtained capture radius for retrograde orbits, $\tilde{R}_{\rm retro}$, given by Equation (\ref{eq:ret-radi}), which is shown by the vertical dotted line.
The semi-major axis and eccentricity decrease due to gas drag (Figure~\ref{fig:orb_ret}(c)), and the Stokes number also decreases gradually.
Then the direction of the orbital motion changes from the retrograde to the prograde direction when ${\rm St} \simeq 1$.
After the planetesimal's eccentricity is sufficiently damped, the planetesimal continues orbital decay, and the decrease of the Stokes number follows the analytic result obtained by Equation (\ref{eq:stokes-pro}).
In the calculations for these typical orbits (Figures~\ref{fig:orb_pro} and \ref{fig:orb_ret}), collision with the central planet is not taken into account, as we mentioned before. 
Thus, in Figure~\ref{fig:r_st}(b), the change of the orbital direction takes place at $\tilde{r} < 10^{-3}$, which is smaller than the value corresponding to the radius of Jupiter.
However, such orbital behavior seems quite common in the vicinity of the planet where the nebular gas density is high (Suetsugu \& Ohtsuki, in preparation).
The radial location corresponding to such a change of the orbital direction can be derived by setting ${\rm St} = 1$ in Equation~(\ref{eq:stokes-retro}) as
\begin{eqnarray} %()
	\tilde{R}_{{\rm St}=1} = (2\zeta)^{1/(\gamma-1)}.
\end{eqnarray}
Our numerical results show that the actual radial distance for the change of the orbital direction is somewhat smaller, and can be approximately given as
\begin{eqnarray} %()
		\tilde{R}_{\rm turn} = \tilde{R}_{{\rm St}=1}/f,
\label{eq:rturn}		
\end{eqnarray}
where $f \sim 2-3$ is a correction factor. 
Planetesimals captured on retrograde orbits spiral into the planet keeping the retrograde orbital direction if $\tilde{R}_{\rm turn} < \tilde{R}_{\rm p}$ ($\tilde{R}_{\rm p}$ is the physical size of the planet scaled by its Hill radius), while they change the orbital direction before hitting the planet if $\tilde{R}_{\rm turn} > \tilde{R}_{\rm p}$.

Figure~\ref{fig:r_st}(c) shows the case of the retrograde capture orbit in the three-dimensional case shown in Figures~\ref{fig:ret_pro}(a) and (b).
The general behavior is similar to the coplanar retrograde case, and the direction of the orbital motion changes as the orbital decay proceeds.
Because of the vertical motion, the planetesimal goes through regions with very low gas density above and below the mid-plane, where the Stokes number takes on very large values.
The minimal values of the Stokes number roughly correspond to the line for ${\rm St}_{\rm retro}$ (Equation~(\ref{eq:stokes-retro})).
The orbital direction changes when ${\rm St} \sim 1$; the actual location for this change depends on various factors, such as the gradual decrease of the orbital inclination and the above-mentioned oscillation of the Stokes number due to the vertical motion.
As in the coplanar case, the Stokes number follows the analytic result obtained by Equation~(\ref{eq:rturn}) after the direction of the orbit is changed and the eccentricity is sufficiently damped so that the assumption of circular orbits in the derivation of the analytic result is verified.

\subsection{Orbital Decay Timescales}

In Section~\ref{sec:long-lived}, we have discussed long-lived orbits under gas drag, which may be important in relation to the origin of irregular satellites of giant planets.
On the other hand, in relation to the supply of solid materials that would become building blocks of regular satellites, evolution of regular orbits in the circumplanetary gas disk is important.
Here, we examine timescales of the evolution of typical prograde and retrograde orbits in the circumplanetary gas disk.
As we have seen in Section~\ref{subsec:orb_evo}, the orbits of planetesimals captured in the prograde direction become nearly circular at the radial location corresponding to the prograde capture radius, while the orbits of planetesimals captured in the retrograde direction become nearly circular prograde orbits approximately at $\tilde{R}_{\rm turn}$ given by Equation (\ref{eq:rturn}).
Afterwards, these planetesimals undergo gradual orbital decay on a timescale given analytically \citep{A76}:
\begin{eqnarray} %()
		\tau_{\rm fall} = \frac{r}{|v_r|} = \frac{T_K}{2\pi\eta} \frac{1+g^2}{2g}.
	\end{eqnarray}	
\noindent
In the above, $v_r$ is the planetesimal's radial velocity, $T_{\rm K}$ is the orbital period for the planet-centered motion, and $g \equiv {\rm St}^{-1}$ can be written as
\begin{eqnarray} %()
		g \equiv \frac{a_{\rm drag}}{u \Omega_K},
	\end{eqnarray}
which can be expressed using the non-dimensional quantities we defined above as
	\begin{equation} %()
		g = \zeta\eta \tilde{r}^{-\gamma+1}.
	\end{equation}
In the above, we have assumed that the orbital eccentricities and inclinations of captured planetesimals are sufficiently small and used $\tilde{u} = \eta \tilde{v}_{\rm K} = \eta \sqrt{3/\tilde{r}}$.
Then, we have
	\begin{eqnarray} \label{eq:tau_fall}
		\tau_{\rm fall}/T_{\rm pla}  = \frac{1}{2\pi\eta} \sqrt{\frac{\tilde{r}^3}{3}} \frac{1+g^2}{2g}.
	\end{eqnarray}
Since $\gamma = 11/4$ in our model, we have $\tau_{\rm fall}/T_{\rm pla} \propto \zeta^{-1}\tilde{r}^{9/4}$ in the limit of $g \ll 1$.

If we substitute the capture radius for prograde orbits, $\tilde{R}_{\rm pro}$, into $\tilde{r}$ in the above expression, we can estimate the timescale of orbital decay after capture in the prograde direction.
Similarly, if we substitute the radial location of the change of the orbital direction $\tilde{R}_{\rm turn}$ for $\tilde{r}$, we obtain the timescale of orbital decay after the retrograde orbits become prograde.
Figure~\ref{fig:timescale}(a) shows the plots of $\tilde{R}_{\rm pro}$ and $\tilde{R}_{\rm turn}$ as a function of the gas drag parameter $\zeta$.
The radius of Jupiter is also shown by the horizontal line for comparison.
These plots show that the change of the orbital direction due to gas drag likely takes place at a radial location interior to $\tilde{R}_{\rm pro}$. 
In the regions with $\tilde{R}_{\rm turn} < \tilde{R}_{\rm J} = 10^{-3}$ (i.e., cases of planetesimals with size larger than $\sim$ 100m in the case of the typical surface density with $\Sigma_d=1$gm$^{-2}$), planetesimals on retrograde orbits collide with the planet before their orbital direction is changed.

Figure~\ref{fig:timescale}(b) shows the orbital decay timescale $\tau_{\rm fall}$ at $\tilde{r} = \tilde{R}_{\rm pro}$ and $\tilde{r} = \tilde{R}_{\rm turn}$, as a function of $\zeta$.
In the case of prograde orbits, the capture radius ($\tilde{R}_{\rm pro}$) increases with increasing $\zeta$, while $\tau_{\rm fall}$ at $\tilde{r} = \tilde{R}_{\rm pro}$ decreases with increasing $\zeta$ as shown in Figure~\ref{fig:timescale}(b). 
With increasing $\zeta$, the Stokes number as well as the orbital decay timescale decreases (i.e., $\tau_{\rm fall}/T_{\rm pla} \propto \zeta^{-1}$). 
On the other hand, with increasing capture radius, the orbital decay timescale becomes longer because captured bodies orbit in the regions with low gas density (i.e., $\tau_{\rm fall}/T_{\rm pla} \propto \tilde{r}^{9/4}$).
Figure~\ref{fig:timescale}(b) shows that the former effect is dominant over the latter effect.
In the case of retrograde capture orbits, the captured body reaches $\tilde{R}_{\rm turn}$ quickly after becoming permanently captured.
In the case of $\tilde{R}_{\rm p} = 10^{-3}$, the change of the orbital direction takes place before hitting the planet if $\zeta \gtrsim 10^{-5}$ (Figure~\ref{fig:timescale}(a)).
The timescale of orbital decay after changing the orbital direction is much shorter than the orbital decay timescale at the prograde capture radius (Figure~\ref{fig:timescale}(b)), which suggests that the lifetime of planetesimals captured in retrograde orbits is much shorter than those captured in prograde orbits, even if the change of the orbital direction is taken into account.
The orbital decay timescale at $\tilde{r} = \tilde{R}_{\rm turn}$ increases with increasing $\zeta$, owing to the effect of the increasing $\tilde{R}_{\rm turn}$ with increasing $\zeta$.

%% file: sec7.tex
\section{CONCLUSIONS AND DISCUSSION}
\label{sec:conclusion}

In the present work, we examined orbital characteristics of planetesimals captured by gas drag from circumplanetary gas disks.
We found that the semi-major axes of planet-centered orbits of planetesimals at the time of permanent capture have an upper limit ($a_{\rm p} \lesssim r_{\rm H}/3$), which can be explained analytically assuming that the capture takes place through energy dissipation in the vicinity of the planet.
In the typical case, capture takes place when planetesimals enter the dense inner part of the disk and undergo strong gas drag, and captured bodies spiral into the planet in a short timescale.
On the other hand, we found that there are certain types of orbits on which captured planetesimals can survive in the circumplanetary disk for a long time even under gas drag.
We found that such long-lived orbits exist both in the prograde and retrograde cases, and they may be important in relation to the origin of irregular satellites of giant planets.
However, in order for such captured bodies to become the irregular satellites we observe today, the circumplanetary gas disk needs to be dissipated before the bodies spiral into the planet.
In the present work, we assumed that the gas drag parameter does not change with time during integration of each orbit.
In order to clarify the processes of the capture and survival of satellites, we will investigate capture of planetesimals by waning circumplanetary gas disks in our separate work \citep{SO16}.

On the other hand, behavior of regular orbits both in the prograde and retrograde directions is important in relation to the supply of solid materials into the circumplanetary gas disk and the formation of regular satellites.
In the typical case of capture in the prograde direction, eccentricities of planet-centered orbits are quite large immediately after the capture.
But captured bodies undergo strong gas drag every time they pass the pericenter, and the semi-major axes and eccentricities decrease rapidly while keeping the pericenter distance nearly unchanged.
After the eccentricity becomes sufficiently small ($e_{\rm p} \lesssim 0.2$), the decrease of the semi-major axis is slowed down.
In the case of the capture in the retrograde direction, the initially large semi-major axes and eccentricities decrease through strong gas drag when captured bodies pass through their pericenter in the dense part of the disk.
When the eccentricity becomes smaller than $\sim 0.2$, their apocenter also enter the regions with rather high gas density, resulting in rapid orbital decay.
In some cases, the direction of orbits changes from the retrograde to the prograde direction due to gas drag. 
In the case of retrograde orbits with low inclinations the direction can change suddenly due to the strong headwind, while the change proceeds rather slowly when planetesimals are initially captured into largely inclined orbits.
Such a change of the orbital direction is likely to take place when the size of planetesimals is smaller than $\sim$10m if the disk surface densities comparable to the so-called gas-starved disk model \citep{CW02} is assumed (Figure~\ref{fig:timescale}(a)).
Thus, it is expected that the fraction of small planetesimals on prograde orbits increases in the vicinity of the planet.

As we have mentioned above, planetesimals have large orbital eccentricities immediately after their capture, regardless of the direction of motion.
Capture of planetesimals into retrograde orbits are more likely to occur owing to the large velocity relative to the gas, but their lifetime in the circumplanetary disk is much shorter than the prograde ones.
Thus, planetesimals captured into prograde orbits would be expected to accumulate in the vicinity of the planet in the circumplanetary disk.
Such distribution of captured bodies would be important for the growth of regular satellites.
Also, dust produced by collision between these captured bodies may also influence the evolution of the satellite systems, and may be observable around extrasolar giant planets.
We will investigate the distribution of captured planetesimals in circumplanetary disks in our subsequent paper. 

\acknowledgments
We thank Takayuki Tanigawa for valuable discussions and advice in the early stage of this work.
This work was supported by JSPS Grants-in-Aid for JSPS Fellows (12J01826) and Scientific Research B (22340125 and 15H03716).
Part of numerical calculations were performed using computer systems at the National Astronomical Observatory of Japan.

%% file: figcap.tex
\clearpage

\begin{figure}[ht]
 \begin{center}
 \epsscale{0.55}
  \plotone{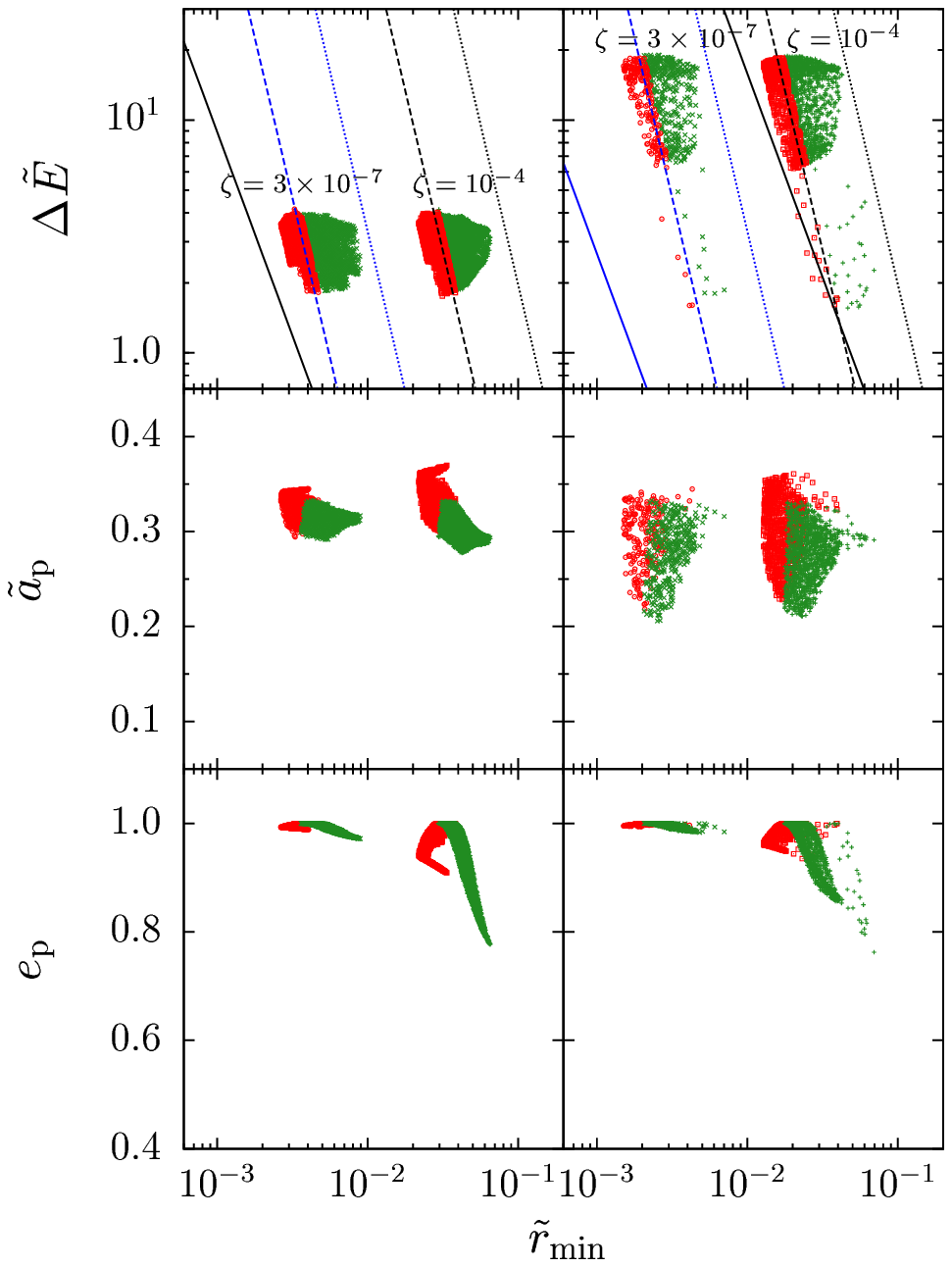}
 \epsscale{1}
  \end{center}
 \caption{
Total amount of energy dissipated before becoming permanently captured ($\Delta \tilde{E}$), and semi-major axes (in units of the planet's Hill radius $r_{\rm H}$) and eccentricities of the planet-centers orbits at the time of permanent capture, as a function of the minimum approach distance to the planet (in units of $r_{\rm H}$).
The cases for planetesimals  captured by a single encounter with the circumplanetary gas disk is shown. 
Red and green marks represent prograde and retrograde capture orbits, respectively.
Solid, dashed, and dotted lines represent $\tilde{R}_{\rm ran}$, $\tilde{R}_{\rm shear,pro}$, and $\tilde{R}_{\rm shear,ret}$, respectively, with black and blue colors showing the cases with $\zeta=10^{-4}$ and $\zeta=3\times10^{-7}$, respectively.
Left and right panels show the case of $e_{\rm H}=0.5$ and $5$, respectively.}
 \label{fig:rmin_all}
\end{figure}

\begin{figure}[ht]
 \begin{center}
  \plottwo{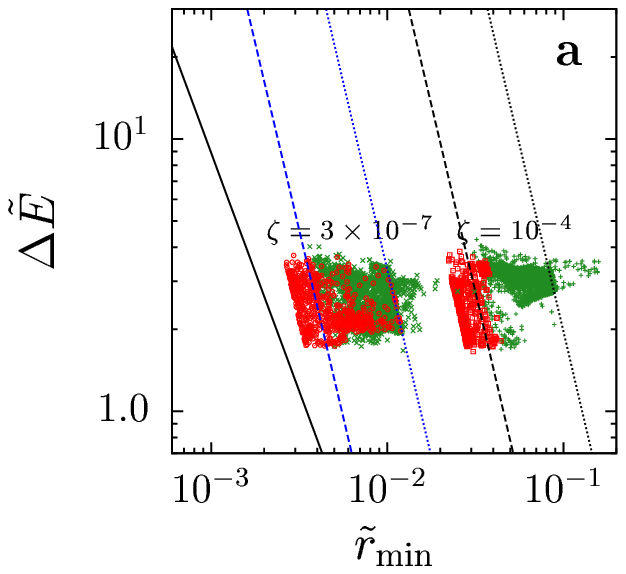}{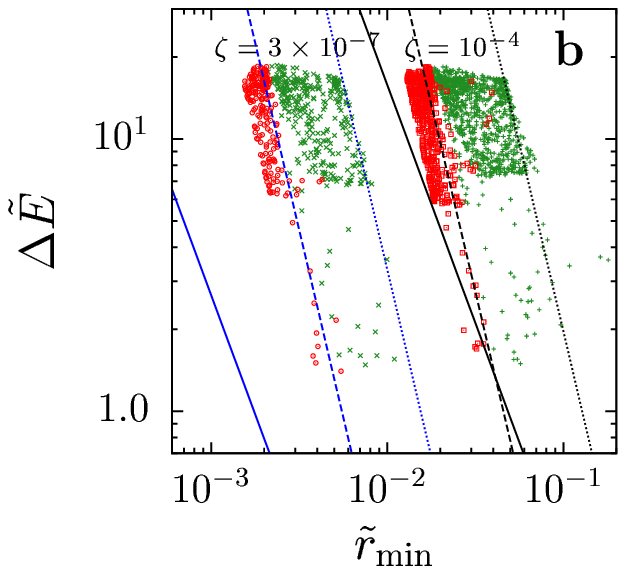}
  \end{center}
 \caption{Total amount of energy dissipated before becoming permanently captured as a function of $\tilde{r}_{\rm min}$ for the cases of capture after multiple encounters with the circumplanetary disk. (a) $e_{\rm H}=0.5$, and (b) 5.}
 \label{fig:rmin_long}
\end{figure}

\begin{figure}
\epsscale{0.7}
\plottwo{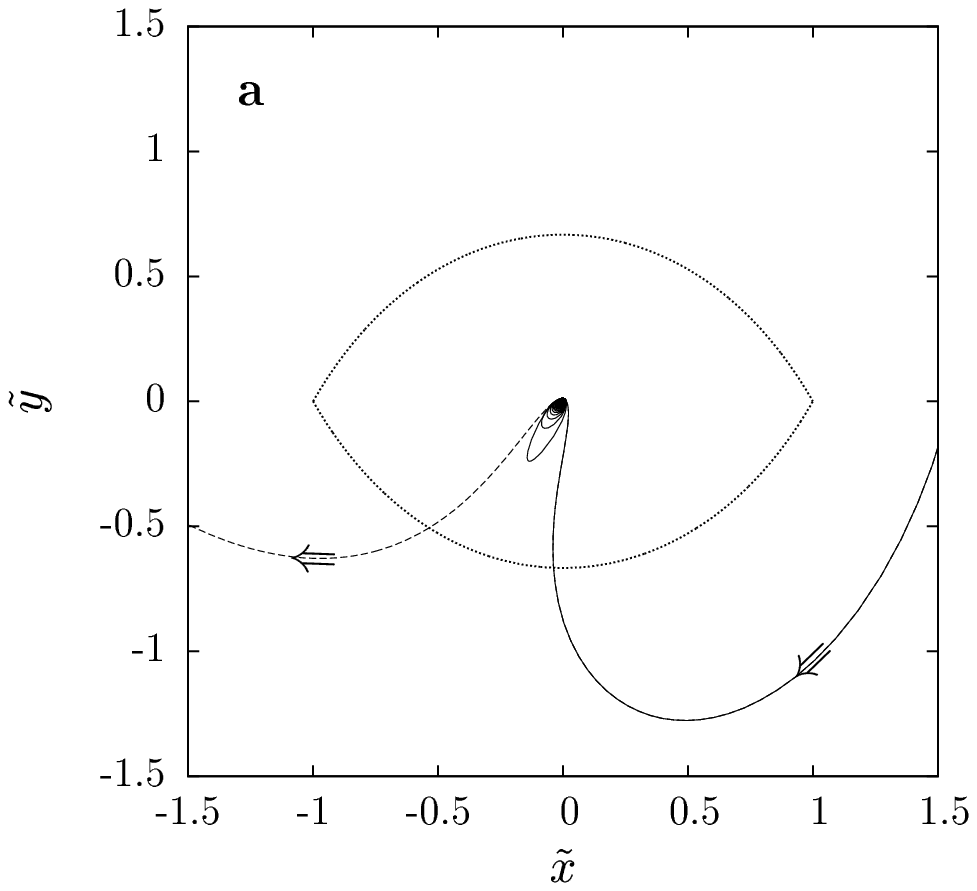}{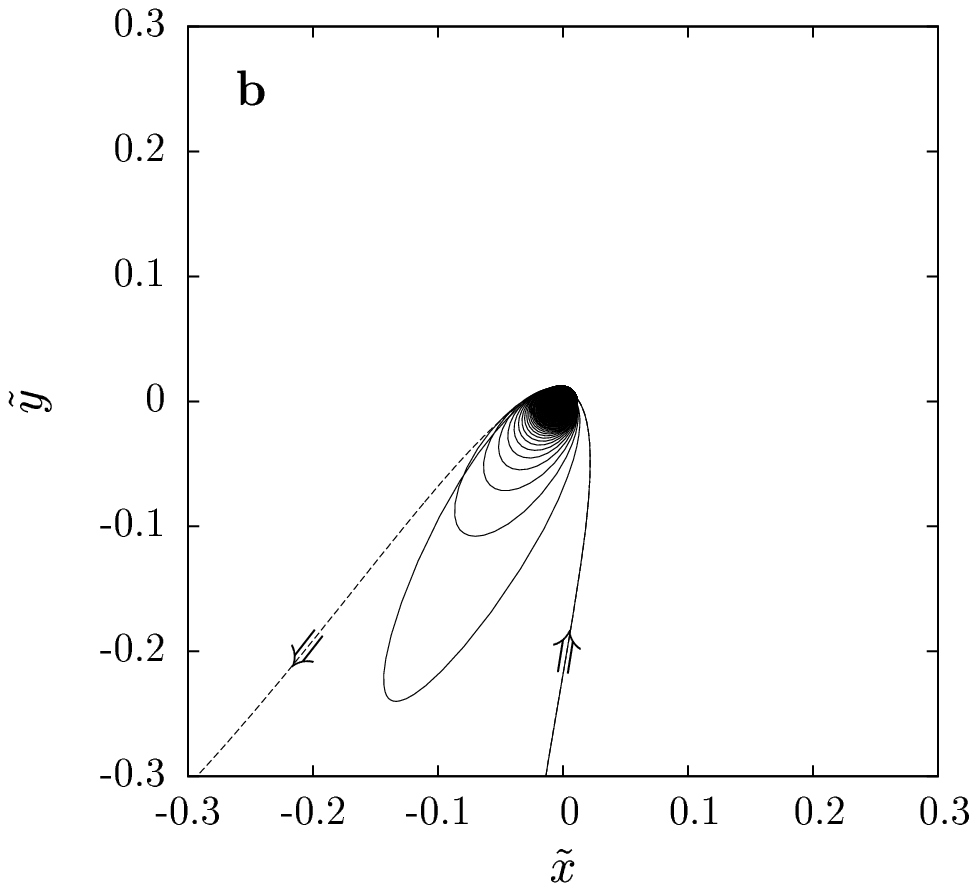}
\epsscale{0.3}
\plotone{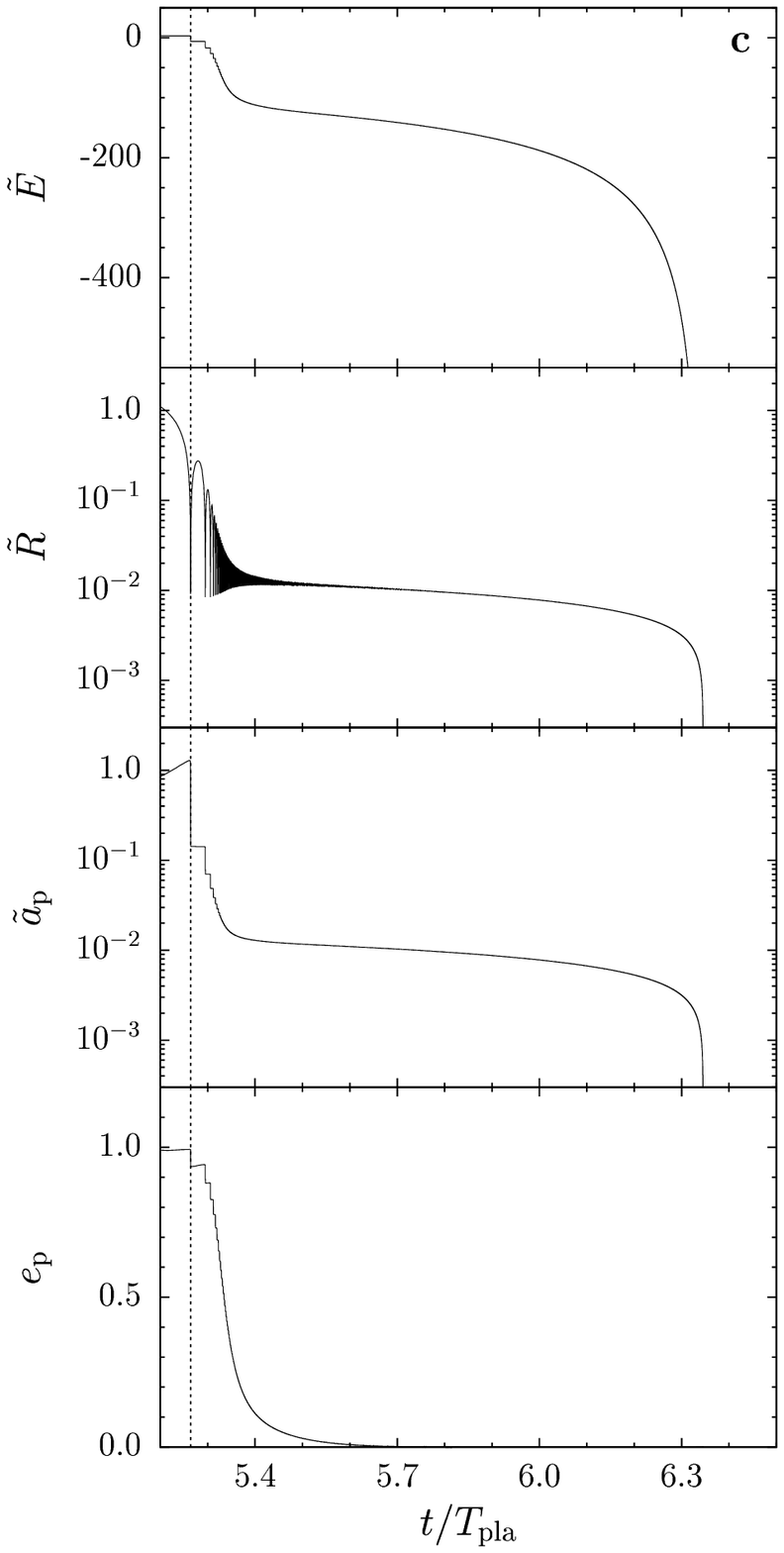}
\epsscale{1}
\caption{
(a) Example of a captured orbit in the prograde direction; initial heliocentric orbital elements are $e_{\rm H}=1$, $i_{\rm H}=0$, $b_{\rm H}=2.24$, and $\tau = 5.22$.
Solid line shows the orbit under gas drag with $\zeta = 10^{-5}$, and dashed line shows the case without gas drag.
The dotted line represents the planet's Hill sphere.
(b) Blow-up of Panel (a).
(c) Change of energy, distance from the planet, semimajor axis and eccentricity of the planet-centered orbit of the captured planetesimal shown in (a) and (b) as a function of time. 
$T_{\rm pla}$ is the orbital period of the planet. 
The vertical dotted lines represent the time when the planetesimal becomes permanently captured.
} \label{fig:orb_pro} 
\end{figure}

\begin{figure}
\epsscale{0.7}
\plottwo{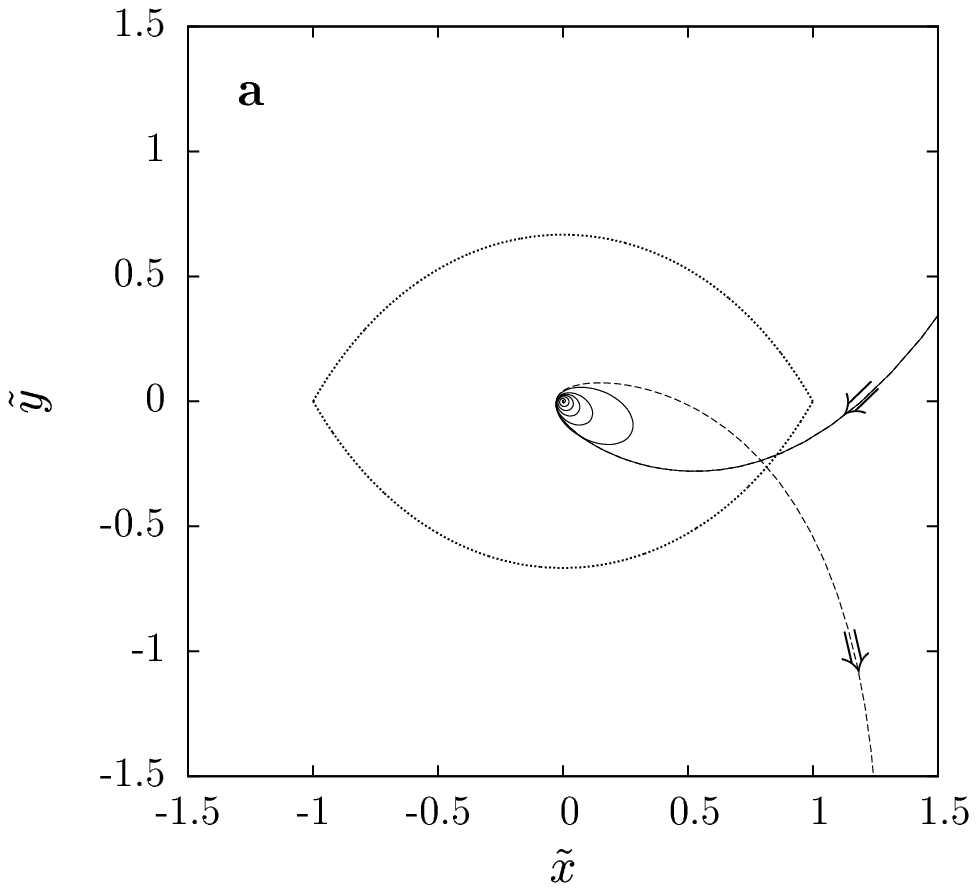}{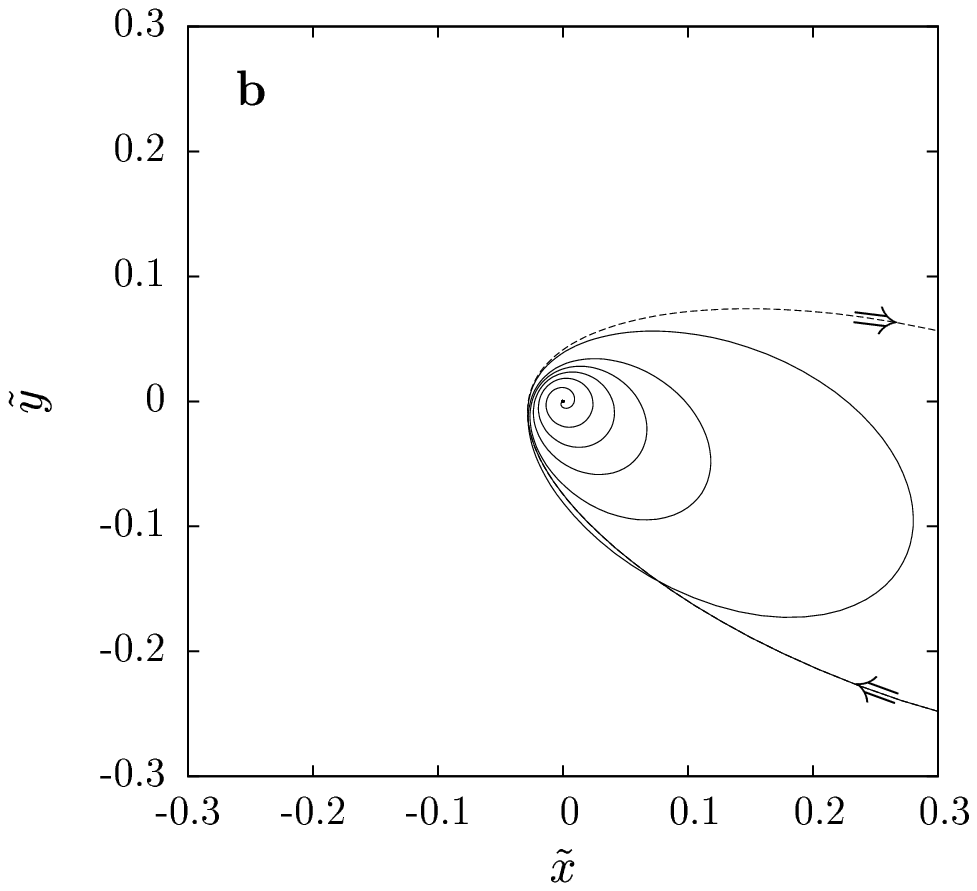}
\epsscale{0.3}
\plotone{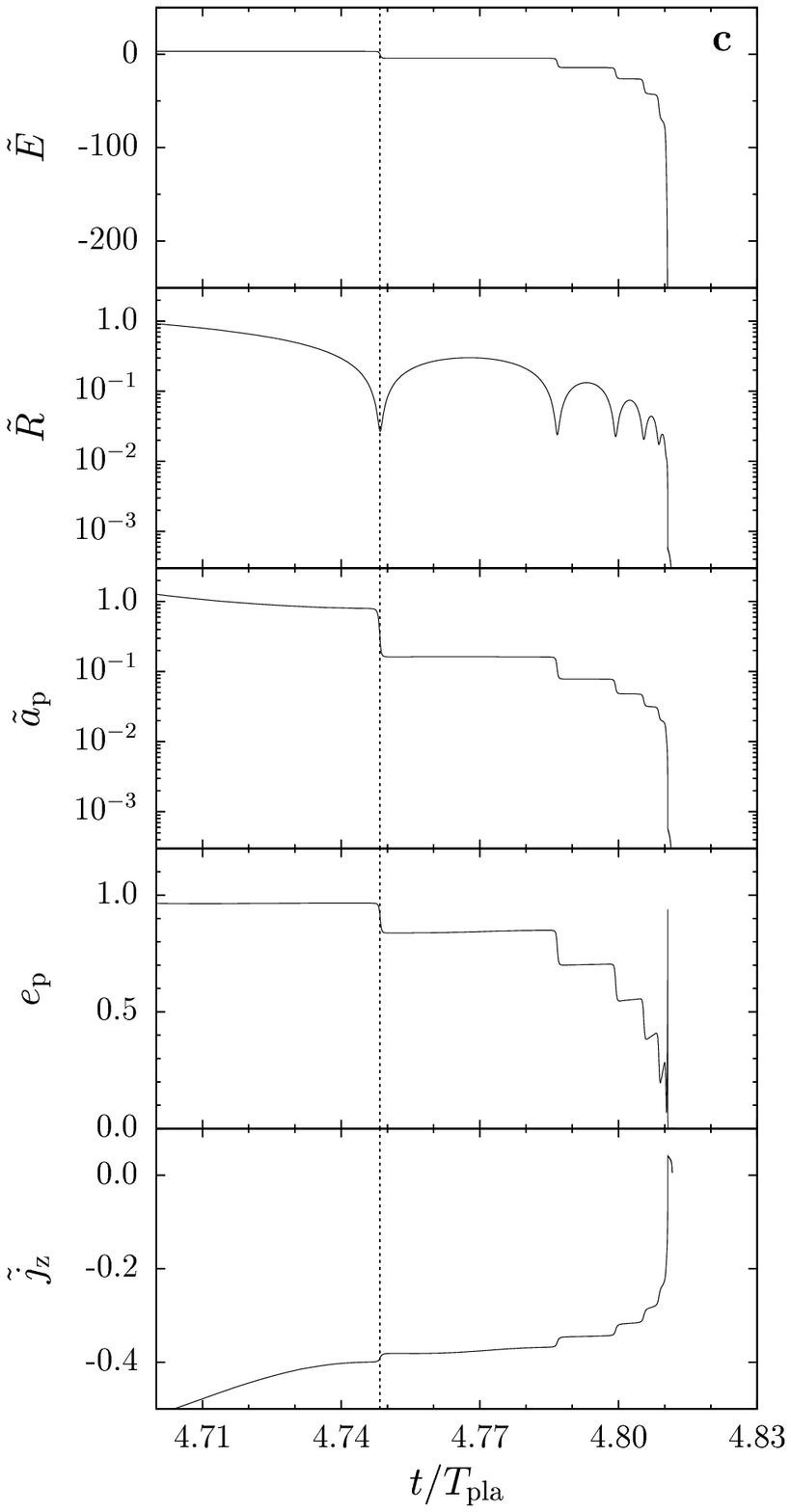}
\epsscale{1}
\caption{Same as Figure~\ref{fig:orb_pro}, but the case of capture in the retrograde direction is shown ($e_{\rm H}=1$, $i_{\rm H}=0$, $b_{\rm H}=2.29$, $\tau = 1.19$).
The right bottom panel shows the time evolution of the $z$-component of the planetesimal's specific orbital angular momentum around the planet in units of $r_{\rm H}^{2}\Omega$.
} \label{fig:orb_ret} 
\end{figure}

\begin{figure}[ht]
 \begin{center}
  \includegraphics[width=120mm]{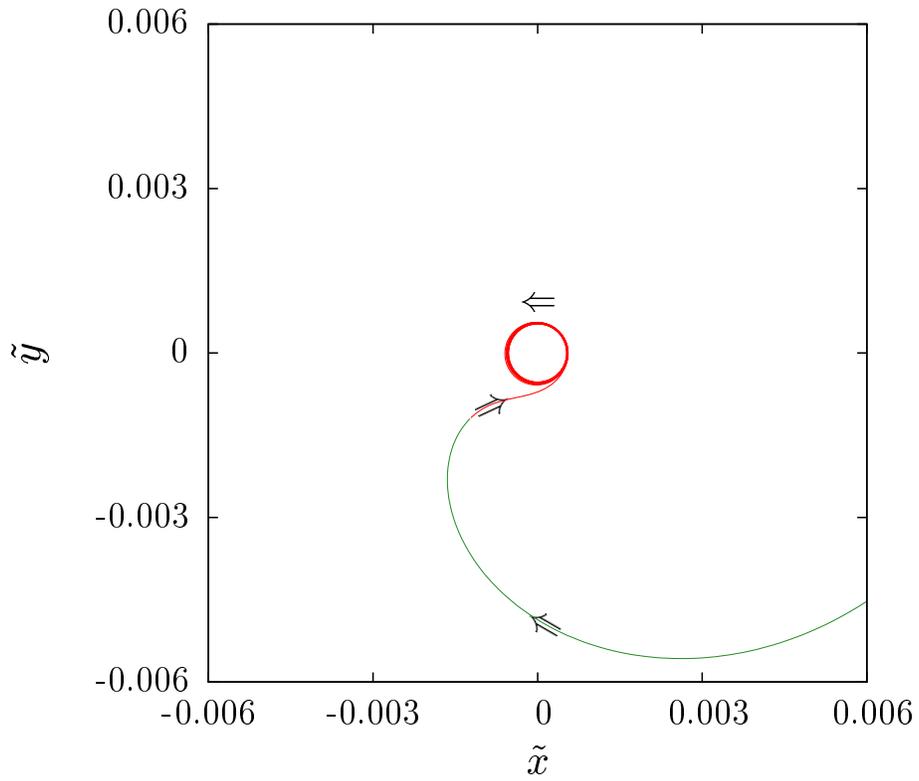}
  \end{center}
 \caption{Blow-up of the orbit shown in Figures~\ref{fig:orb_ret}(a) and \ref{fig:orb_ret}(b).
Captured planetesimal changes the direction of the orbital motion from the retrograde (green) to the prograde (red) direction.}
 \label{fig:orb_ret_up}
\end{figure}

\begin{figure}
  \plottwo{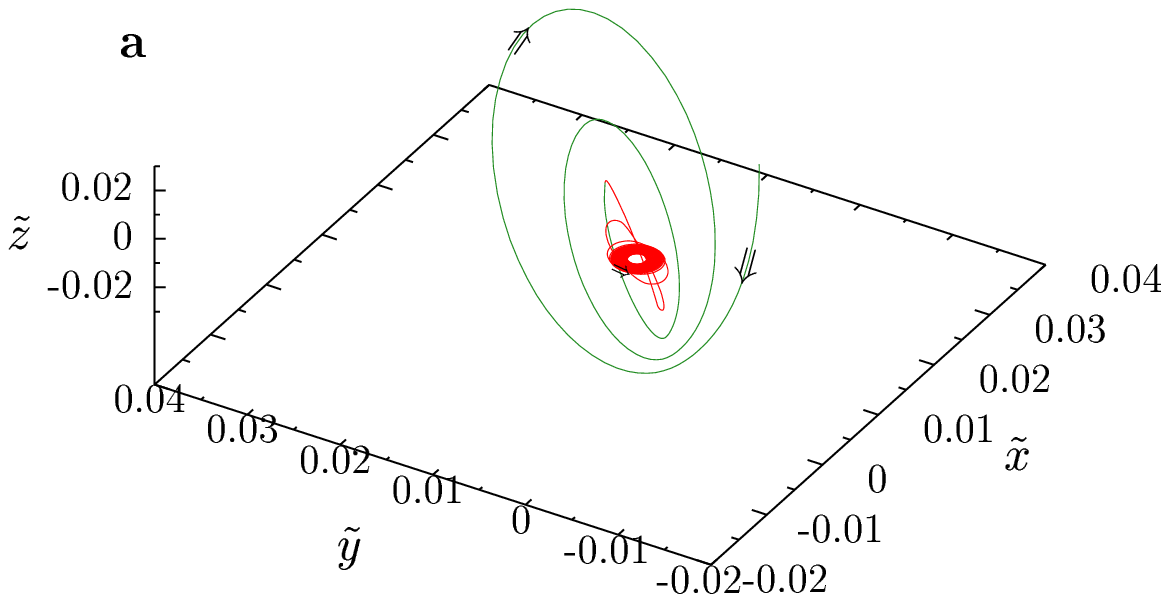}{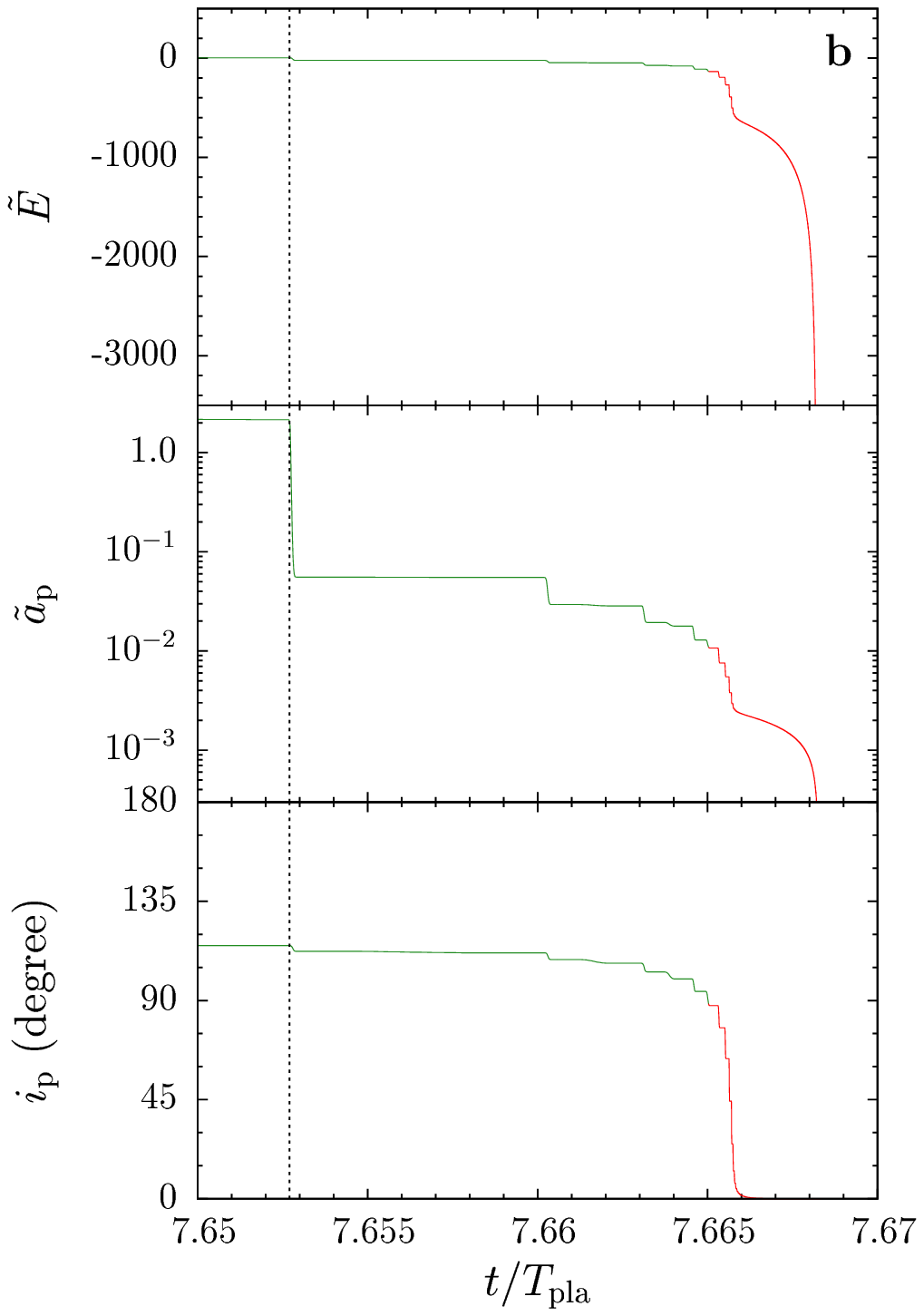}
\caption{
(a) Example of an off-plane capture orbit in the retrograde direction,  which turns to the prograde direction due to gas drag ($\zeta=10^{-4}$, $e_{\rm H}=2i_{\rm H}=1$,  $b_{\rm H}=1.6625$, $\tau=2.3499$, $\omega=4.8506$, and $\tilde{E}=4.089$).
(b)  Change of energy,  semimajor axis, and inclination of the orbit shown in Panel (a) as a function of time.
The vertical dotted lines represent the time when the planetesimal becomes permanently captured.
Red and green lines shows the prograde and retrograde phases, respectively.
} \label{fig:ret_pro} 
\end{figure}

\begin{figure}
\epsscale{0.45}
\plotone{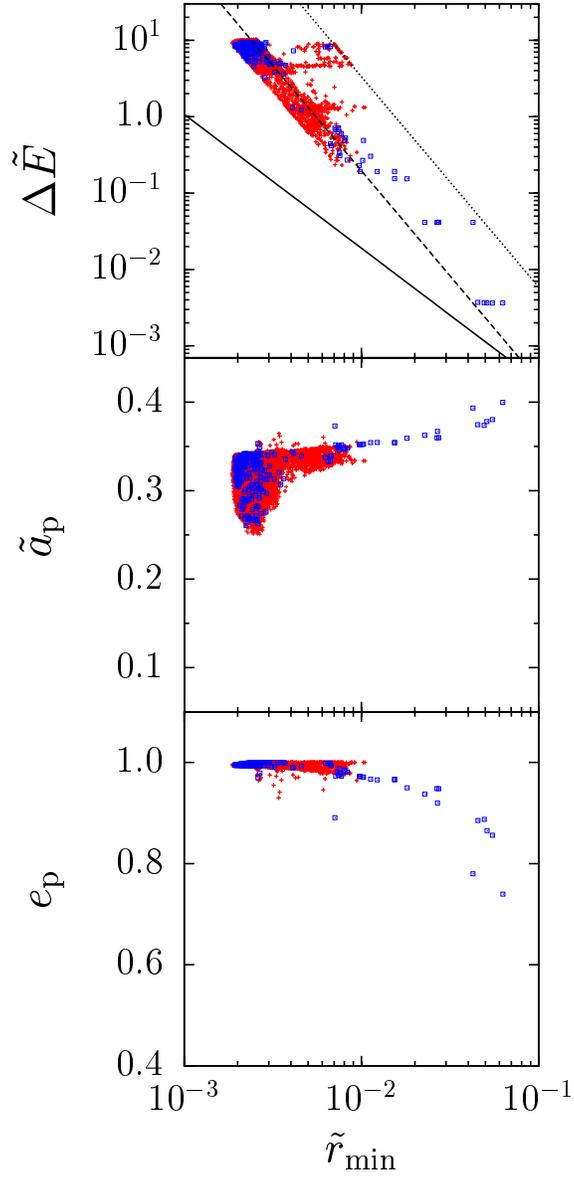}
\epsscale{1.0}
\caption{
Total amount of dissipated energy before becoming permanently captured and orbital elements of the planet-centered orbits at the time of capture as a function of $\tilde{r}_{\rm min}$ for the case of capture in the prograde direction from heliocentric orbits with $e_{\rm H}=3.16$ ($\zeta=3\times10^{-7}, i_{\rm H}=0$).
Red crosses show the case of capture by a single encounter, 
while blue squares show the case of capture after multiple encounters.} 
\label{fig:rmin_pro}
\end{figure}

\begin{figure}
\plottwo{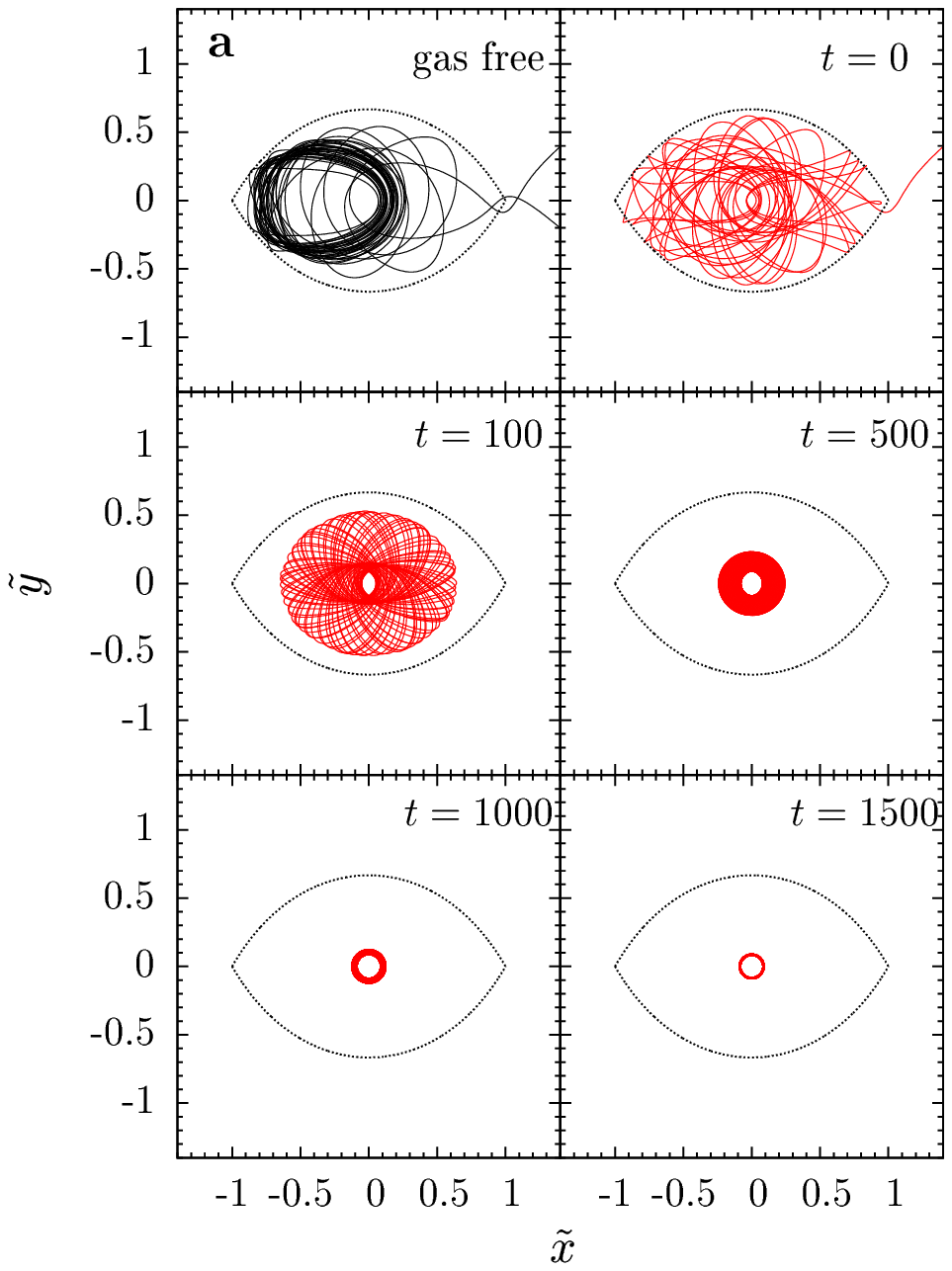}{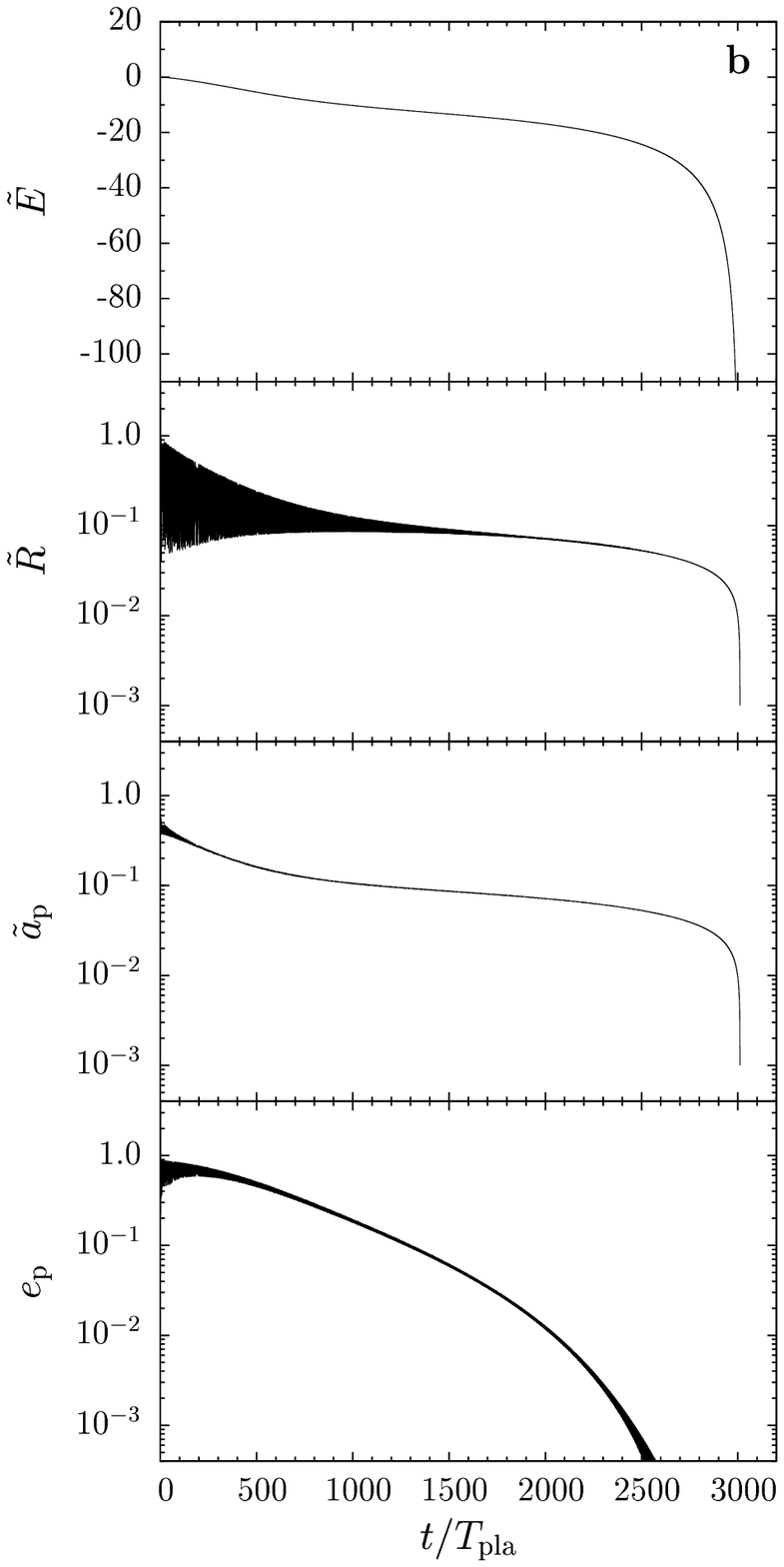}
\caption{
(a) Evolution of a long-lived prograde captured orbit ($\zeta=5\times10^{-7}$, $e_{\rm H}=3$, $i_{\rm H}=0$, $b_{\rm H}=4.890196$, $\tau=0.000295$, $\tilde{E}=0.03058$).
Top-left panel shows the temporary capture orbit in the gas-free environment.
Other panels show orbital evolution due to gas drag for a period of 10$T_{\rm pla}$ from $t/T_{\rm pla}=0, 100, 500, 1000$, and $1500$, respectively.
(b) Time variation of several quantities for the orbit under gas drag shown in Panel (a).
} \label{fig:orb_pro_long} 
\end{figure}

\begin{figure}
%\plottwo{f40.eps}{f41.eps}
\plottwo{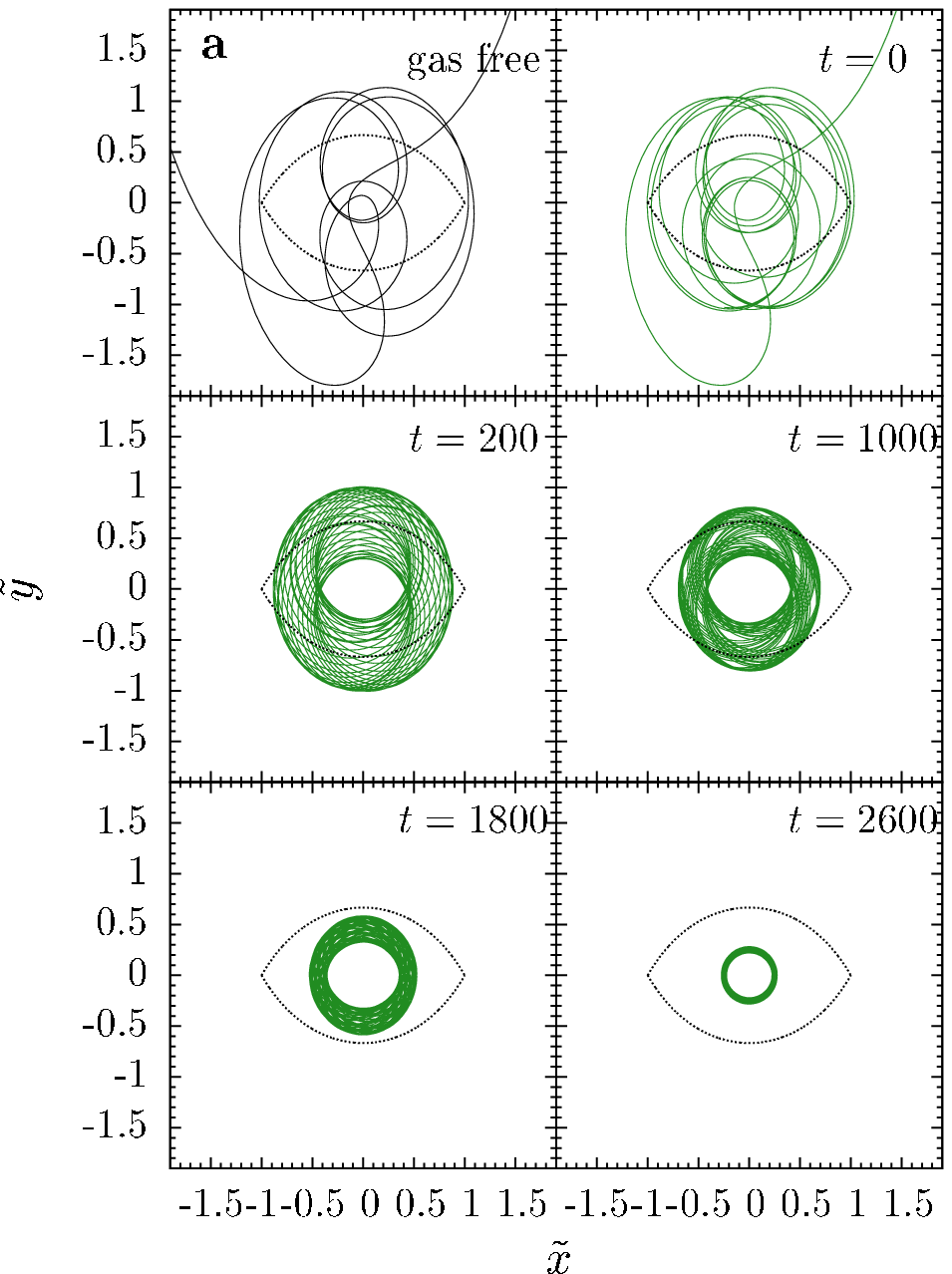}{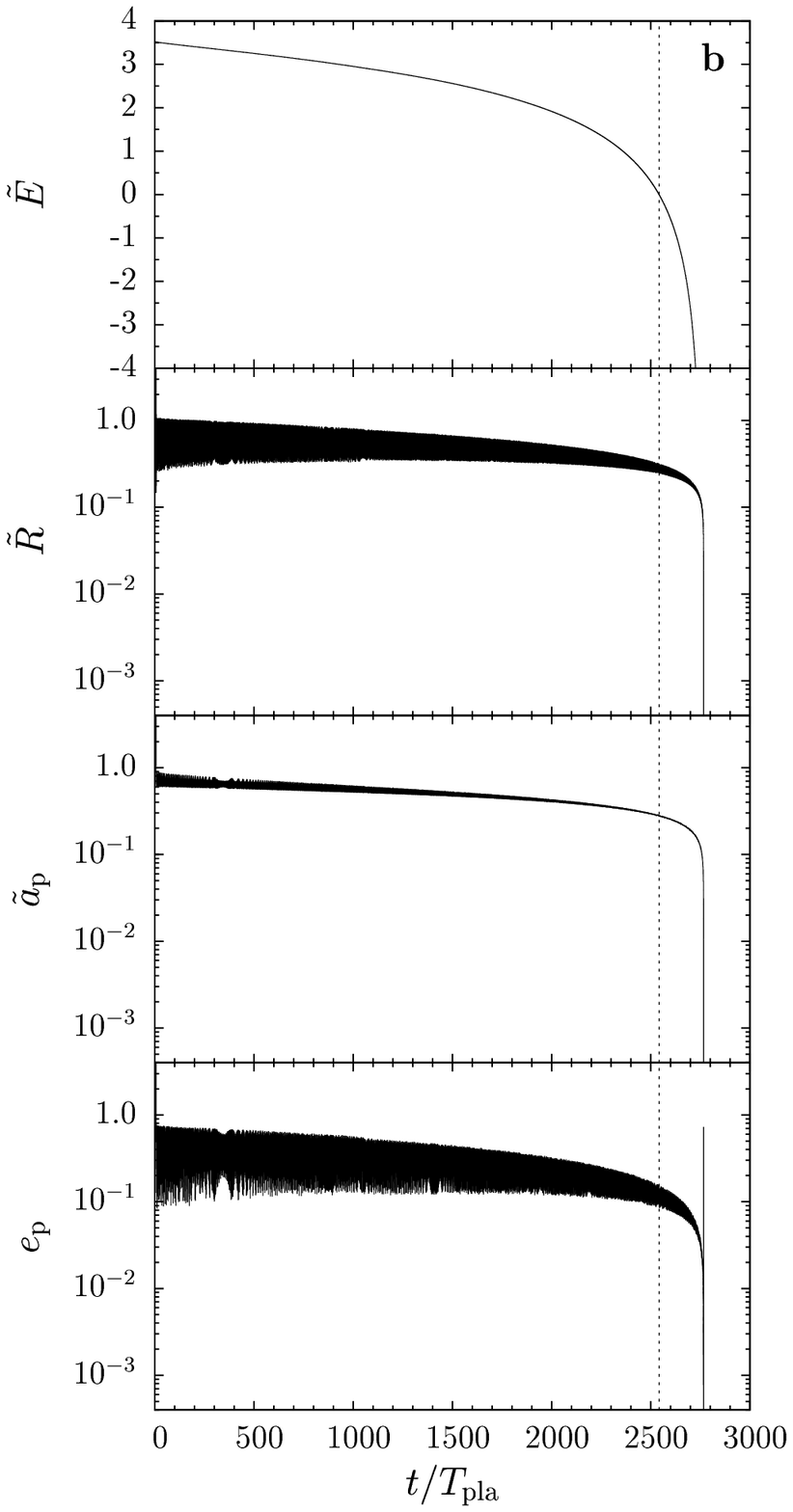}
\caption{
(a) Evolution of a long-lived retrograde captured orbit ($\zeta=3\times10^{-7}$, $e_{\rm H}=0.56234$, $i_{\rm H}=e_{\rm H}/2$, $b_{\rm H}=1.775$, $\tau=2.38761$, $\omega=4.75008$, $\tilde{E}=3.714$).
Top-left panel shows the temporary capture orbit in the gas-free environment.
Other panels show orbital evolution due to gas drag for a period of 10$T_{\rm pla}$ from $t/T_{\rm pla}=0, 200, 1000, 1800$, and $2600$, respectively.
(b) Time variation of several quantities for the orbit under gas drag shown in Panel (a).
} \label{fig:orb_ret_long} 
\end{figure}

\begin{figure}
\epsscale{0.45}
\plotone{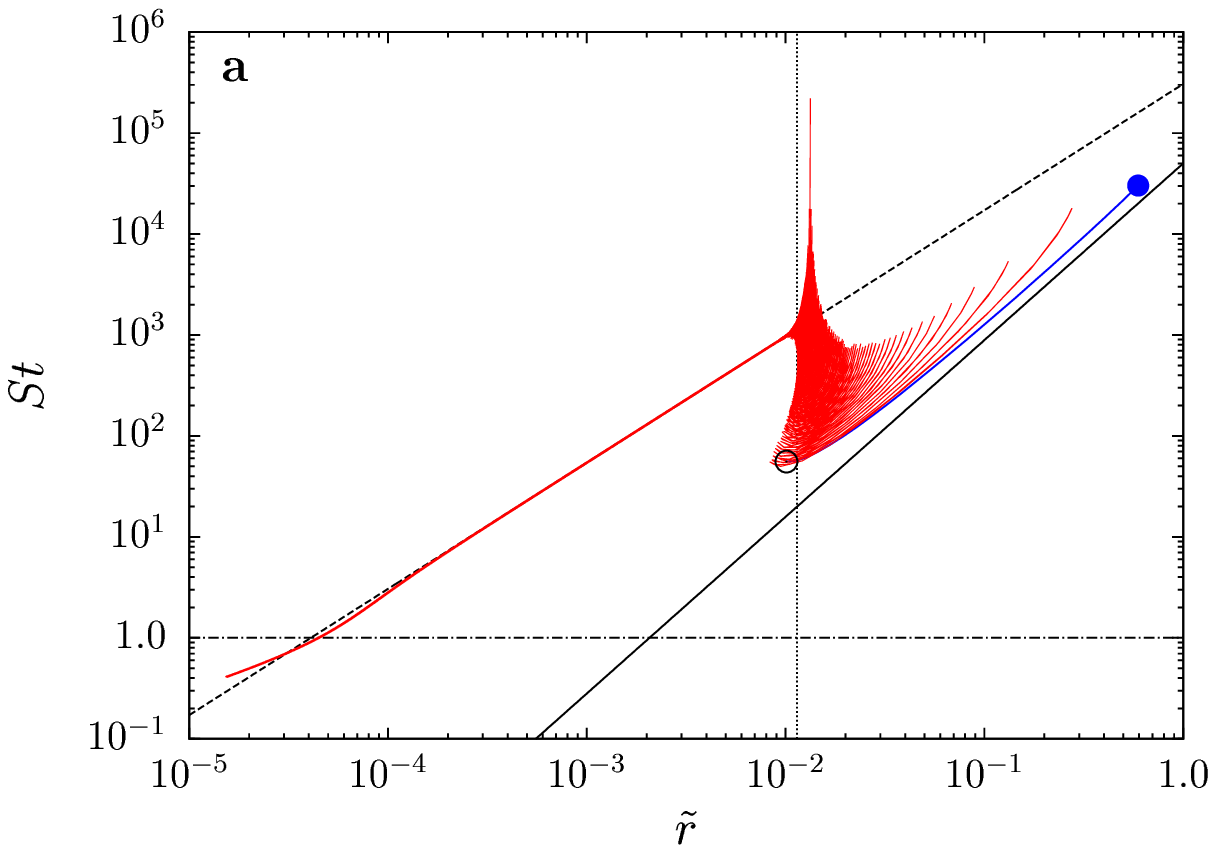}
\plotone{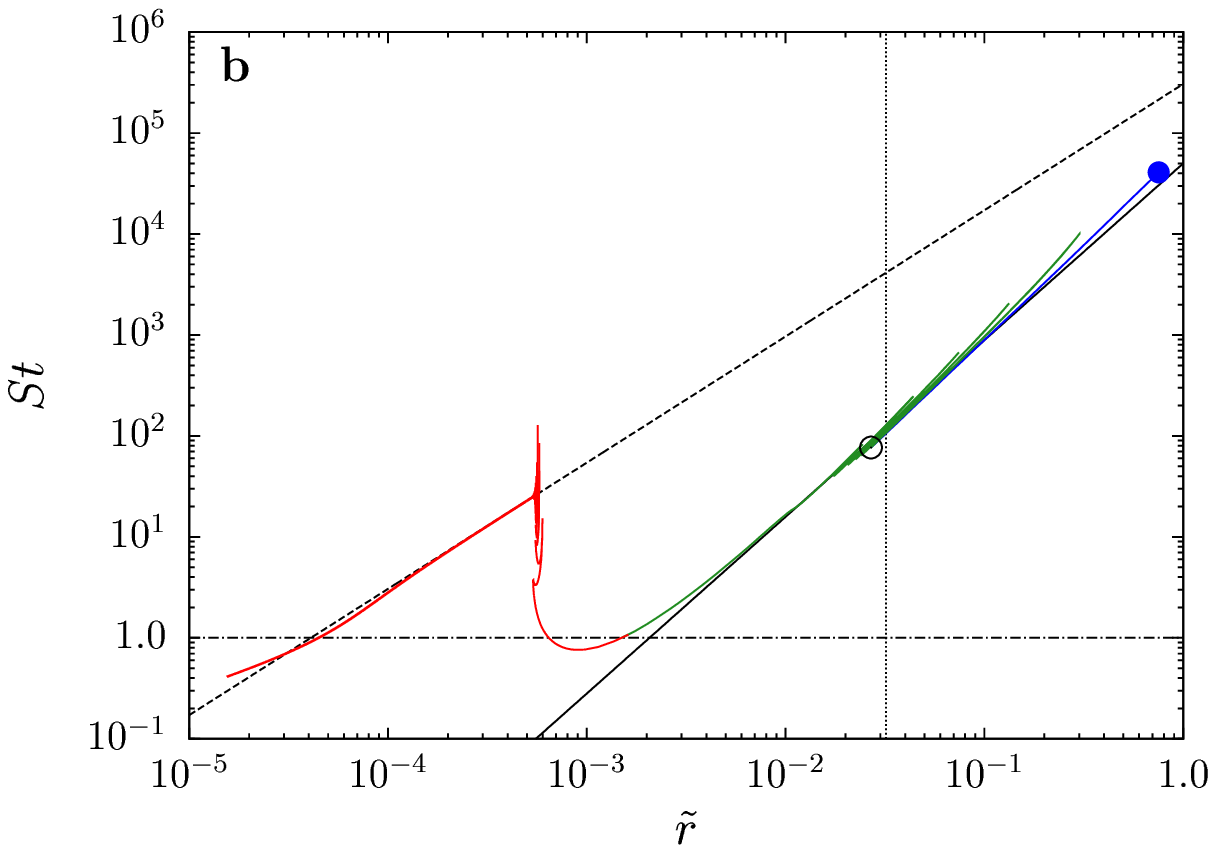}
\plotone{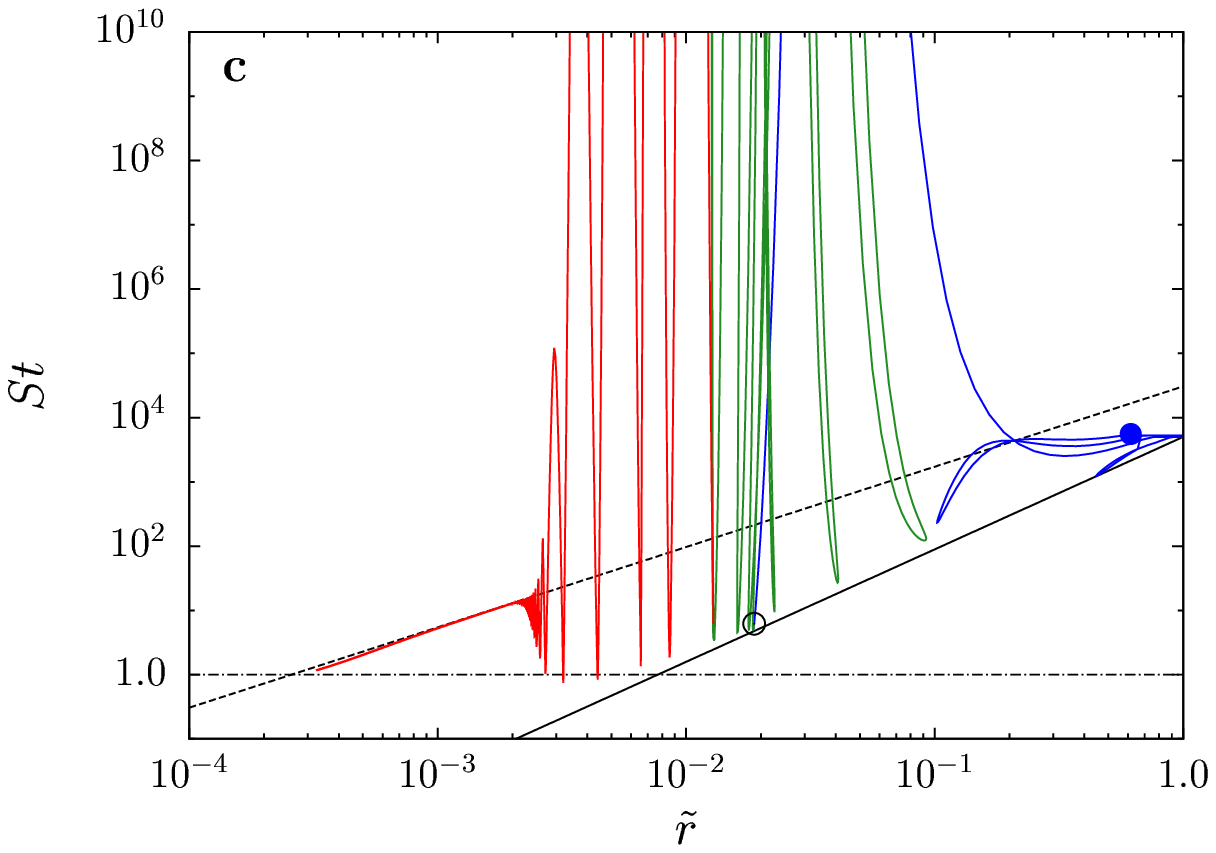}
\epsscale{1.0}
\caption{
Change of the Stokes number for various capture orbits as a function of radial distance from the planet in the mid-plane of the circumplanetary disk. 
Blue lines represent the phase of temporary capture, and red and green lines show the prograde and retrograde phases of permanently captured orbits.
Black dashed and solid diagonal lines represent ${\rm St}_{\rm pro}$ (Eq.(\ref{eq:stokes-pro})) and ${\rm St}_{\rm retro}$(Eq.(\ref{eq:stokes-retro})), respectively, and the horizontal dot-dashed line shows ${\rm St}=1$.
Solid circles show the beginning of the plot, and the open circles show the radial location where the planetesimal becomes permanently captured.
(a) Case for the prograde capture orbit shown in Figure~\ref{fig:orb_pro}. 
(b) Case for the retrograde capture orbit shown in Figure~\ref{fig:orb_ret}. 
(c) Case for the retrograde capture orbit shown in Figure~\ref{fig:ret_pro}.
The vertical dotted lines in Panels (a) and (b) represent the analytically-obtained capture radius for each case.
} \label{fig:r_st} 
\end{figure}

\begin{figure}
\plottwo{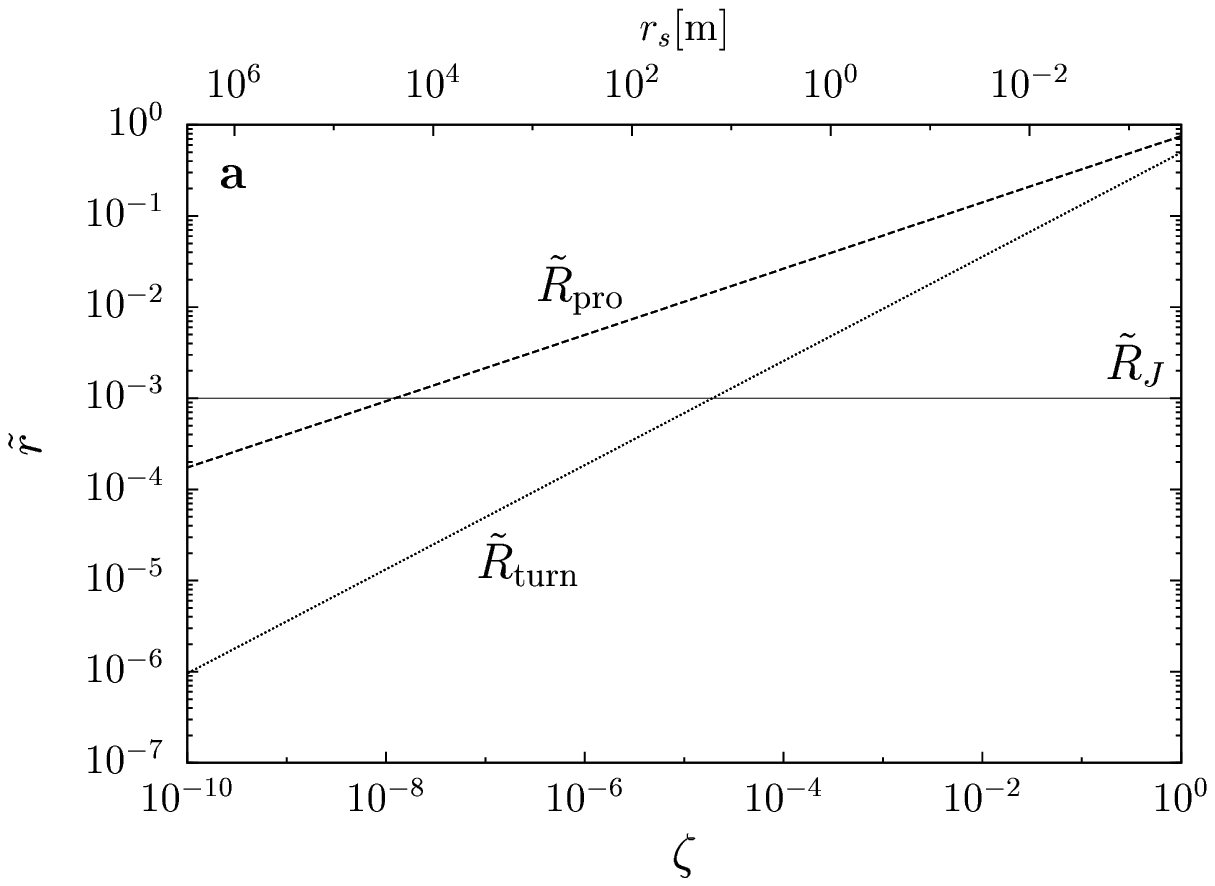}{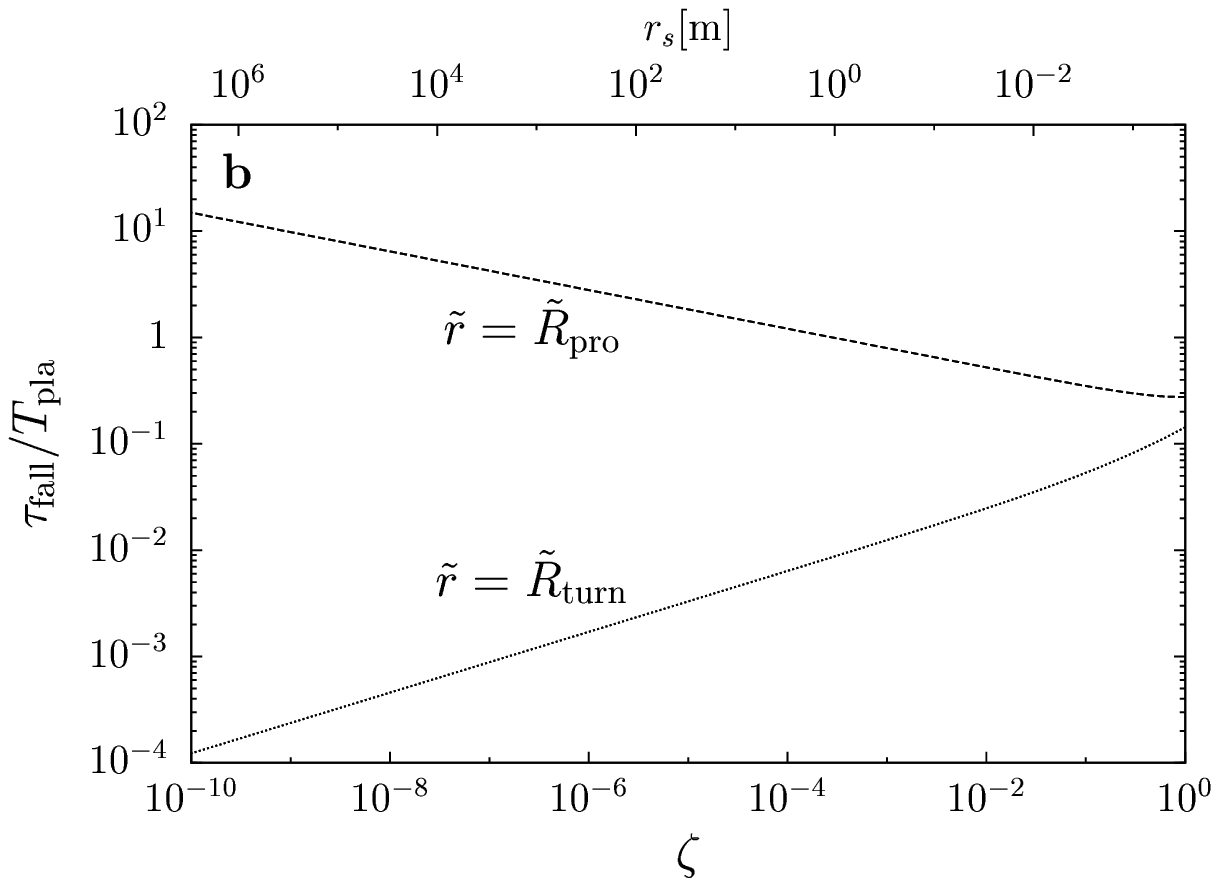}
\caption{
(a) Prograde capture radius (dashed line) and the radial distance for the change of the orbital direction (dotted line; $f=3$) as a function of $\zeta$.
Horizontal line represents the physical size of Jupiter.
In the case of $g \ll 1$ (or $\zeta \ll 1$), $\tilde{R}_{\rm pro} \propto \zeta^{4/11}$ and $\tilde{R}_{\rm turn} \propto \zeta^{4/7}$.
(b) Timescale of the decay of prograde orbits at $\tilde{r}=\tilde{R}_{\rm pro}$ (dashed lines) and $\tilde{r}=\tilde{R}_{\rm turn}$ (dotted line) as a function of $\zeta$.
In the case of $g \ll 1$ (or $\zeta \ll 1$), $\tau_{\rm fall}/T \propto \zeta^{-2/11}$ and $\propto \zeta^{2/7}$ for $\tilde{r}=\tilde{R}_{\rm pro}$ and $\tilde{r}=\tilde{R}_{\rm turn}$, respectively.
Upper horizontal axis shows sizes of planetesimals for a circumplanetary disk with $\Sigma_{d}=1$gcm$^{-2}$ in Equation~(\ref{eq:zeta}).
} \label{fig:timescale}
\end{figure}